\documentclass[acmtog,screen,nonacm]{acmart}

\AtBeginDocument{%
  }
\setcopyright{none}
\renewcommand\footnotetextcopyrightpermission[1]{}
\usepackage{graphicx}
\usepackage{multirow}
\usepackage{makecell}
\usepackage{rotating}
\usepackage{tabularx}
\usepackage{wrapfig}
\newlength{\savedintextsep}
\newlength{\savedcolumnsep}
\AtBeginDocument{\setlength{\savedintextsep}{\intextsep}\setlength{\savedcolumnsep}{\columnsep}}
\begin{document}

\title{OffsetAxis: UDF Mesh Reconstruction via Offset-Volume Medial Axis Extraction}

\author{Qijia Huang}
\email{qijia.huang@unistra.fr}
\affiliation{%
  \institution{ICube, Université de Strasbourg, CNRS}
  \city{Strasbourg}
  \country{France}
}
\author{Pierre Kraemer}
\email{kraemer@unistra.fr}
\affiliation{%
  \institution{ICube, Université de Strasbourg, CNRS}
  \city{Strasbourg}
  \country{France}
}
\author{Dominique Bechmann}
\email{bechmann@unistra.fr}
\affiliation{%
  \institution{ICube, Université de Strasbourg, CNRS}
  \city{Strasbourg}
  \country{France}
}

\begin{abstract}
Unsigned distance fields (UDFs) offer broader modeling capabilities than signed distance fields (SDFs), enabling the representation of shapes with open boundaries, non-manifold structures or mixed curve and surface parts.
However, extracting coherent meshes from UDFs is fundamentally harder, as classical grid-based iso-surfacing techniques are not applicable since they require a way to distinguish the inside from the outside of the shape.
We introduce OffsetAxis, a new UDF reconstruction pipeline that supports open, non-manifold, and curve-like geometries.
Our key insight is that the 0-level set extraction problem can be restated as the extraction of the medial axis of the $\alpha$-offset volume of the UDF.
This formulation unlocks mature medial-axis machinery that naturally supports boundaries, non-manifold junctions and curves.
To avoid the biases of grid-based techniques, we sample the $\alpha$-offset surface using ray casting and optimize medial balls inside the offset volume with an efficient variant of Variational Medial Axis Sampling.
The final mesh is recovered by taking the dual of the connectivity of the medial ball clusters, producing structurally coherent reconstructions for a wide range of topologies.
The robustness and versatility of the approach allow it to handle imperfect distance fields, including neural UDFs trained on noisy inputs, the Quasi-Medial Distance Field (Q-MDF), as well as distances computed directly on triangle soups or point clouds.
Extensive experiments demonstrate that our method produces more faithful mesh reconstruction and better alignment with the underlying shape structure than prior techniques.
\end{abstract}


\begin{teaserfigure}
  \includegraphics[width=\textwidth]{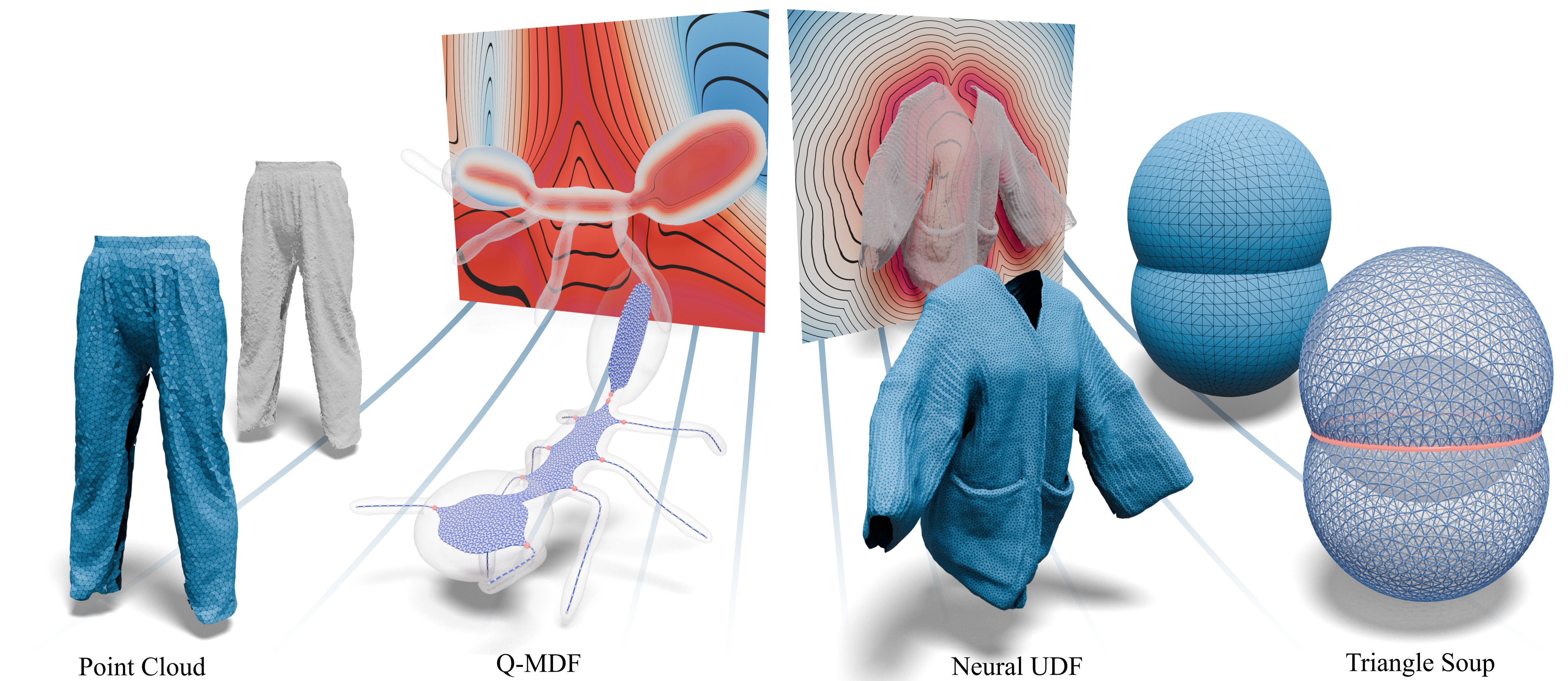}
  \caption{Starting from a function providing the distance from any point in space to the shape, our method can extract structurally coherent meshes from various input types: (a) raw point clouds, (b) learned quasi-medial distance fields, (c) learned unsigned distance fields, or (d) triangle soups. The reconstructed meshes can be open or closed, manifold or non-manifold, and may contain both surface and curve parts.}
  \label{fig:teaser}
\end{teaserfigure}

\maketitle

\section{Introduction}

Implicit functions have been used to represent 3D geometry for decades, both in classical geometry processing and in more recent learning-based pipelines.
Among them, signed distance fields (SDFs) have become a \textit{de facto} standard: the sign induces a globally consistent inside/outside notion, which in turn makes Marching Cubes (MC) iso-surfacing techniques or sphere-tracing algorithms reliable in practice.
However, requiring a globally consistent orientation also restricts the scope of SDFs: open surfaces and non-manifold structures cannot be represented as they do not admit a well-defined global sign.
In contrast, unsigned distance fields (UDFs) do not require global inside/outside labeling and therefore naturally apply to open and non-manifold geometries, a property which has recently attracted considerable attention.
Nevertheless, the lack of sign information makes mesh extraction from UDFs substantially more challenging, as classical iso-surfacing techniques such as MC are no longer directly applicable.

A common workaround is to retrofit MC by reintroducing a notion of sign or orientation, either by propagating local directions (e.g. from estimated normals or UDF gradients) or by predicting cube-wise sign patterns with a learned classifier.
In practice, enforcing global consistency is difficult, and these heuristics often break around noise, thin structures, and non-manifold junctions.
Another approach exploits a double-covering pipeline \cite{Hou2023DCUDF} in which an $\alpha$-offset surface of the UDF is first extracted using MC before being projected back onto the target 0-level set.
However, in cases like non-manifold objects, this method outputs a two-sheet double cover with duplicated layers and incorrect connectivity.

In this paper, we introduce OffsetAxis, a UDF mesh reconstruction pipeline that produces single-sheet, structurally coherent meshes across challenging settings, including open boundaries, non-manifold junctions and curve-like geometry.
Our key observation is that reconstructing the 0-level set of a UDF can be reformulated as the extraction of the medial axis of an $\alpha$-offset volume of the UDF.
Crucially, our approach completely avoids constructing an explicit representation of the $\alpha$-offset surface.
Instead, we bypass the biases of grid-based techniques by employing a point-based sampling strategy that combines raycasting \cite{raycastSampling} with Poisson disk sampling.
We then leverage the recently proposed Variational Medial Axis Sampling (VMAS) \cite{VMAS} pipeline to optimize a set of medial balls inside the $\alpha$-offset volume.
As an additional contribution, we improve this method by introducing a linear regularization term that enables the replacement of the non-linear sphere optimization with a single closed-form step, eliminating inner iterations and substantially accelerating the process.
Finally, we extract the mesh as the dual of the connectivity of the clusters of surface samples induced by the optimized balls.
Furthermore, we pay particular attention to the very common setting where the input UDF is imperfect.
This includes neural representation of UDFs trained on noisy inputs, as well as the Quasi-Medial Distance Field (Q-MDF) \cite{QMDF}, obtained by the joint learning of the SDF and Medial Field (MF) \cite{rebain2021dmf} of a given shape, resulting in a field that is not an actual distance function.

Through extensive qualitative and quantitative evaluations on synthetic open and non-manifold test cases, as well as diverse real-world datasets, we show that OffsetAxis reconstructs non-manifold junctions, open boundaries, and sharp features more faithfully than prior UDF meshing pipelines.

\section{Related work}

Early methods such as Neural Distance Fields (NDF) \cite{chibane2020ndf} use the UDF gradients to extract a dense point cloud and subsequently triangulate these points using the ball-pivoting algorithm.
While capable of recovering the surface, this two-step process is quite inefficient and generally produces poor surface quality.

A major category of methods focuses on artificially recovering sign information or pseudo-signs at the corners of a spatial grid, effectively reducing the problem to standard MC extraction.
Methods like MeshUDF \cite{guillard2022meshudf} and CAP-UDF \cite{zhou2024capudf} assign these pseudo-signs by analyzing the relative orientations of local UDF gradients.
MeshUDF pairs this with a breadth-first surface-following heuristic to enforce global sign consistency.
Concurrently, GeoUDF \cite{ren2023geoudf} introduced an Edge-based Marching Cubes algorithm that bypasses corner signs altogether, instead testing for surface intersections directly on the grid edges.
Because hand-crafted gradient rules are highly sensitive to noise and often introduce topological artifacts, recent works utilize neural networks for sign inference.
GIFS \cite{ye2022gifs} predicts surface-edge interactions using a network, while NSD-UDF \cite{stella2024nsdudf} trains a model to predict pseudo-signs at voxel corners based on local UDF values and gradients, ensuring compatibility with existing SDF meshing pipelines.
A critical limitation of single-pass grid methods is that they perform poorly at high resolutions, where neural UDFs become increasingly noisy and ambiguous, leading to holes and missing surfaces in the results.
To resolve this, Iterative Networks \cite{stella2025high} propose a recursive approach that repeatedly refines cell configurations.
By spatially propagating UDF values, gradients, and previously estimated signs from neighboring cells across multiple passes, this method stabilizes extraction and successfully recovers highly detailed geometry.
\begin{figure}[t]
    \centering
    \includegraphics[width=\columnwidth]{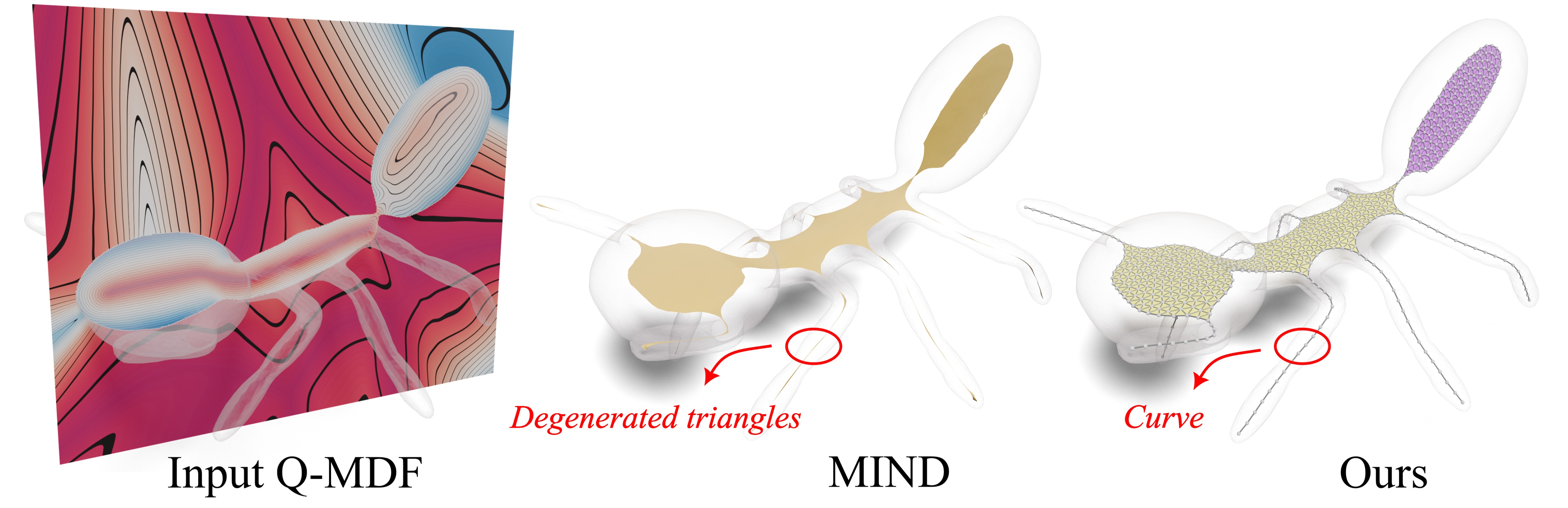}
    \caption{This example illustrates the reconstruction of a medial axis encoded by a Q-MDF learned from a given input surface mesh. Our method produces a clean mixed-dimensional mesh composed of surface and curve parts, whereas MIND, like the other grid-based methods, outputs a triangle mesh with degenerate geometry.}
    \label{fig:bug_curve}
\end{figure}
As an alternative to estimating pseudo-signs, double-covering-based methods start with the extraction of an inflated proxy mesh.
DCUDF \cite{Hou2023DCUDF} extracts an $\alpha$-level iso-surface which forms a watertight, double-layered manifold, and optimizes a loss function to project this boundary onto the true zero-level set, followed by a minimum-cut graph algorithm to separate the layers.
Building upon this, DCUDF2 \cite{chen2025dcudf2} addresses the original method's tendency to over-smooth high-curvature details by introducing self-adaptive weights, an accuracy-aware loss function, and dynamic topology correction.
Despite these improvements, double-covering methods are fundamentally limited by their reliance on deforming a manifold proxy and are inherently incapable of modeling true non-manifold geometries, often outputting duplicated overlapping layers.

Dual methods attempt to bypass pseudo-signs by placing vertices directly inside grid cells, theoretically capturing sharp edges and complex junctions.
DMUDF (DualMesh-UDF) \cite{zhang2023dmudf} optimizes a quadratic error function to place vertices, pruning empty cells using a distance threshold.
Similarly, NDC \cite{chen2022ndc} relies on learned priors via a 3D CNN to predict dual vertex locations.
However, these approaches suffer from severe topological connectivity issues.
If the optimized vertex falls outside its bounding cell, the algorithm either forces erroneous connectivity or discards the vertex, leaving noticeable holes in the mesh.
Furthermore, these dual formulations frequently introduce unintended non-manifold artifacts in strictly manifold regions.

\begin{figure*}[ht]
    \centering
    \includegraphics[width=\linewidth]{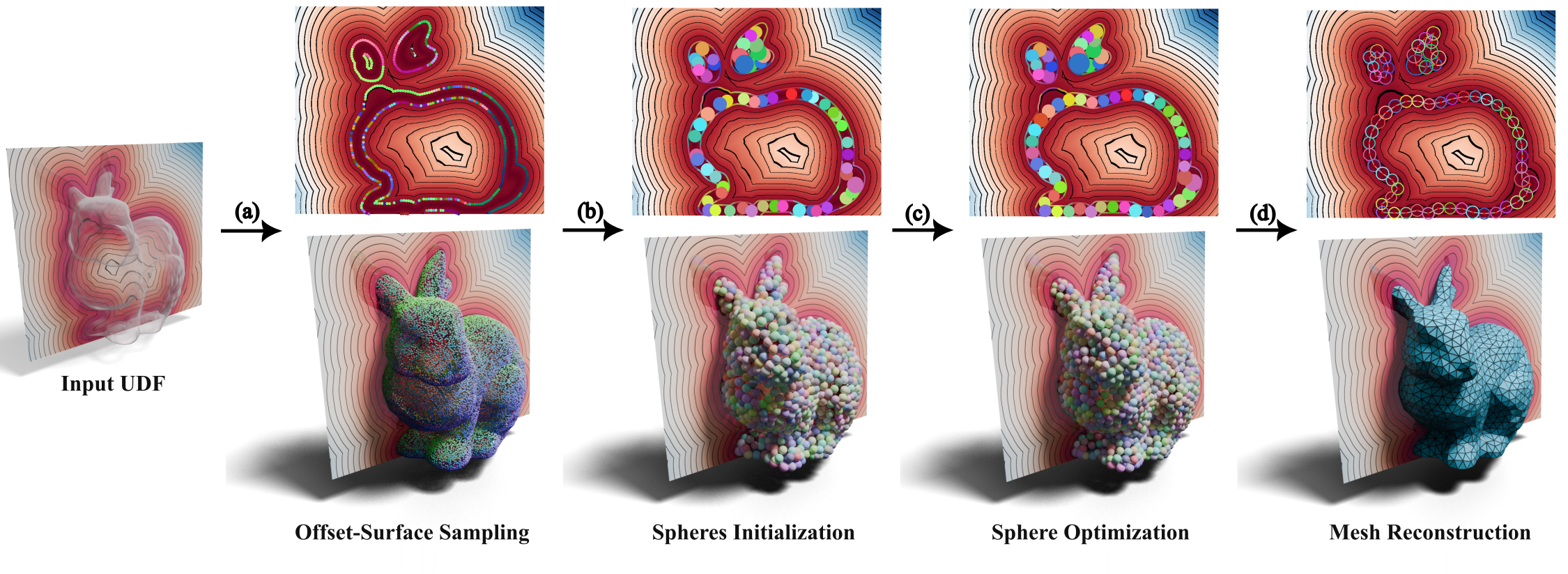}
    \caption{\textbf{Overview of the method:} (a) Starting from a given UDF, its $\alpha$-offset surface is sampled with a raycasting-based approach. (b) These samples are projected onto the medial axis of the offset volume, producing a set of maximally inscribed spheres, among which a subset is selected using a covering approach. (c) These spheres are then optimized with a simple variational formulation relating each sphere to a cluster of surface samples. (d) The final mesh is extracted by connecting the sphere centers using the connectivity of their clusters.}
    \label{fig:overview}
\end{figure*}

Recent advancements aim to extract geometrically accurate, non-manifold structures by circumventing the assumptions of standard grid-based and double-covering techniques.
For instance, MIND \cite{chen2025mind} computes a global multi-labeled field over a voxel grid directly from the UDF, partitioning space to assign unique labels to each side of a non-manifold shape.
The target surface is then extracted as the material interface between these regions using a multi-label MC algorithm.
Despite its general efficacy, its reliance on inherent voxel grid operations, such as morphological erosion, can inadvertently remove structural details.
The labeling strategy also fundamentally breaks down in the presence of non-orientable configurations that naturally occur in complex non-manifold geometries.
Finally, similar to prior methods, MIND is restricted to outputting triangle meshes and cannot natively reconstruct curve-like geometric features as illustrated in Figure \ref{fig:bug_curve}.

\section{Method}
\label{sec:method}

The input of our method is a 3D shape represented by a queryable unsigned distance field (UDF): $\phi:\mathbb{R}^3\rightarrow\mathbb{R}_{\ge 0}$, which encodes, for any query point $x\in\mathbb{R}^3$, its unsigned distance $\phi(x)$ to the shape.
The shape is thus the 0-level set of this field: $S=\{x\in\mathbb{R}^3\mid \phi(x)=0\}$.
As an output, our method produces a mesh composed of triangles for surface regions and segments for curve-like branches.
Our starting point is that UDF mesh reconstruction can be expressed as a medial-axis extraction problem.

Let $\alpha>0$ be an offset parameter.
For each point $p\in S$, consider the closed ball of radius $\alpha$ centered at $p$,
\[
B_\alpha(p)=\{x\in\mathbb{R}^3\mid \|x-p\|\le \alpha\},
\]
and define the $\alpha$-offset volume as the union of these balls,
\[
\Omega_\alpha \;=\; \bigcup_{p\in S} B_\alpha(p).
\]
Equivalently, this offset volume is the $\alpha$-sublevel set of the UDF, and its boundary is the corresponding $\alpha$-level set:
\[
\Omega_\alpha \;=\; \{\,x\in\mathbb{R}^3 \mid \phi(x)\le \alpha\,\},
\qquad
\partial\Omega_\alpha \;=\; \{\,x\in\mathbb{R}^3 \mid \phi(x)= \alpha\,\}.
\]

In the general case, the medial axis of $\Omega_\alpha$ and the target shape $S$ are not exactly the same object.
However, if the value of $\alpha$ is chosen to be smaller than the weak feature size of $S$ \cite{chazal2005WFS}, then $\Omega_\alpha$ and in turn its medial axis are homotopy equivalent to $S$.
The weak feature size is defined as the minimum distance between $S$ and the critical points of the distance function.
Such critical points are characterized by being enclosed in the convex hull of its closest points on $S$.
When the shape is only available through an implicit representation, computing its weak feature size is a problem on its own.
A small $\alpha$ value will lead to a more precise correspondence between $S$ and the medial axis of $\Omega_\alpha$.
However, most UDF representations, and notably many trained neural representations, are not exact, can be non-zero on the shape $S$ or even not be a real distance function.
Too small $\alpha$ values will then often lead to the creation of holes and disconnections in the reconstructed object.
Our goal is to propose a versatile and practical method that works on real-world, potentially unreliable data.
In practice, we suppose that the user has some prior knowledge about $S$ and its representation and is able to choose $\alpha$ to target the desired scale of geometric and structural features..
Controlling this value can also determine the size of the features that should be captured or filled by the reconstruction.

In our setting, the distance function $\phi$ is only available through pointwise queries and $\partial\Omega_\alpha$ is not given explicitly.
A straightforward way to evaluate it would be to mesh $\partial\Omega_\alpha$ via a marching cubes approach.
However, this requires a dense regular discretization of $\phi$, which leads to many extraneous function evaluations and introduces significant memory cost as well as orientation and grid-dependent sampling artifacts that we seek to avoid.
Additionally, the subsequent steps of our pipeline do not need $\partial\Omega_\alpha$ to be available as a surface mesh, in order to extract its medial mesh.
Therefore, we adopt a sampling strategy based on ray intersections and Poisson sampling, yielding uniformly distributed points on $\partial\Omega_\alpha$ (Section \ref{sec:sampling}).
The gradient of $\phi$ should provide normals to these point samples, but as many fields are not reliable, we estimate local neighborhoods to derive normals.
Each of these samples is then associated to a candidate medial sphere, among which an initial subset is selected using a covering strategy (Section \ref{sec:initialization}).
The medial spheres are then optimized following an SQEM-based fitting metric that measures the discrepancy between each sphere and the local geometry of its assigned samples.
This objective is minimized by alternately updating the sample-to-sphere clustering and the sphere parameters, which progressively reduces the fitting error and regularizes the sphere distribution (Section \ref{sec:sphere_fitting_clustering}).
Finally, the medial mesh is obtained by connecting the spheres following the dual of their clusters connectivity, and a final cleanup procedure produces the final output mesh (Section \ref{sec:mesh_reconstruction}).
\subsection{Offset-surface sampling}
\label{sec:sampling}
\setlength{\columnsep}{8pt}
\setlength{\intextsep}{0pt}
\begin{wrapfigure}{r}{0.35\columnwidth}
    \centering
    \includegraphics[width=0.33\columnwidth]{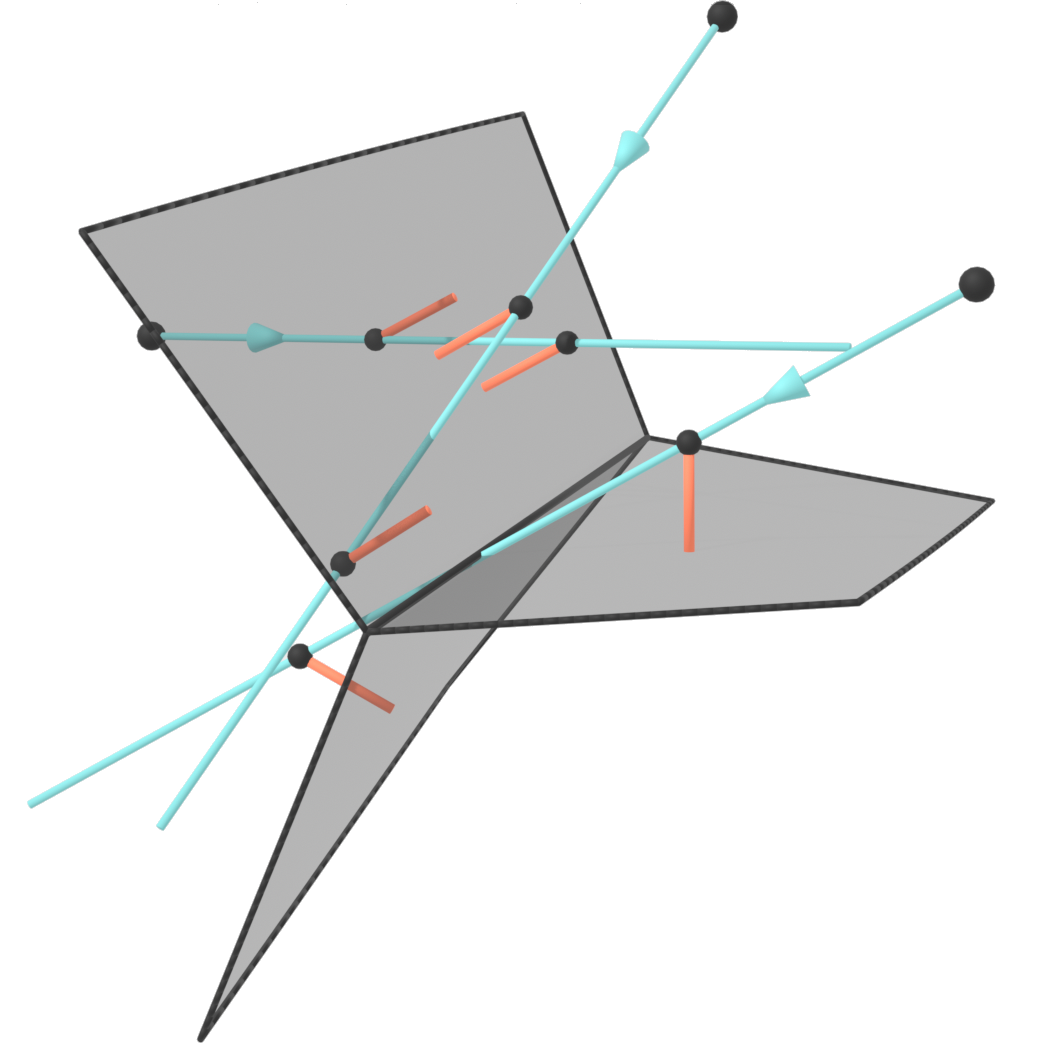}
    \includegraphics[width=0.33\columnwidth]{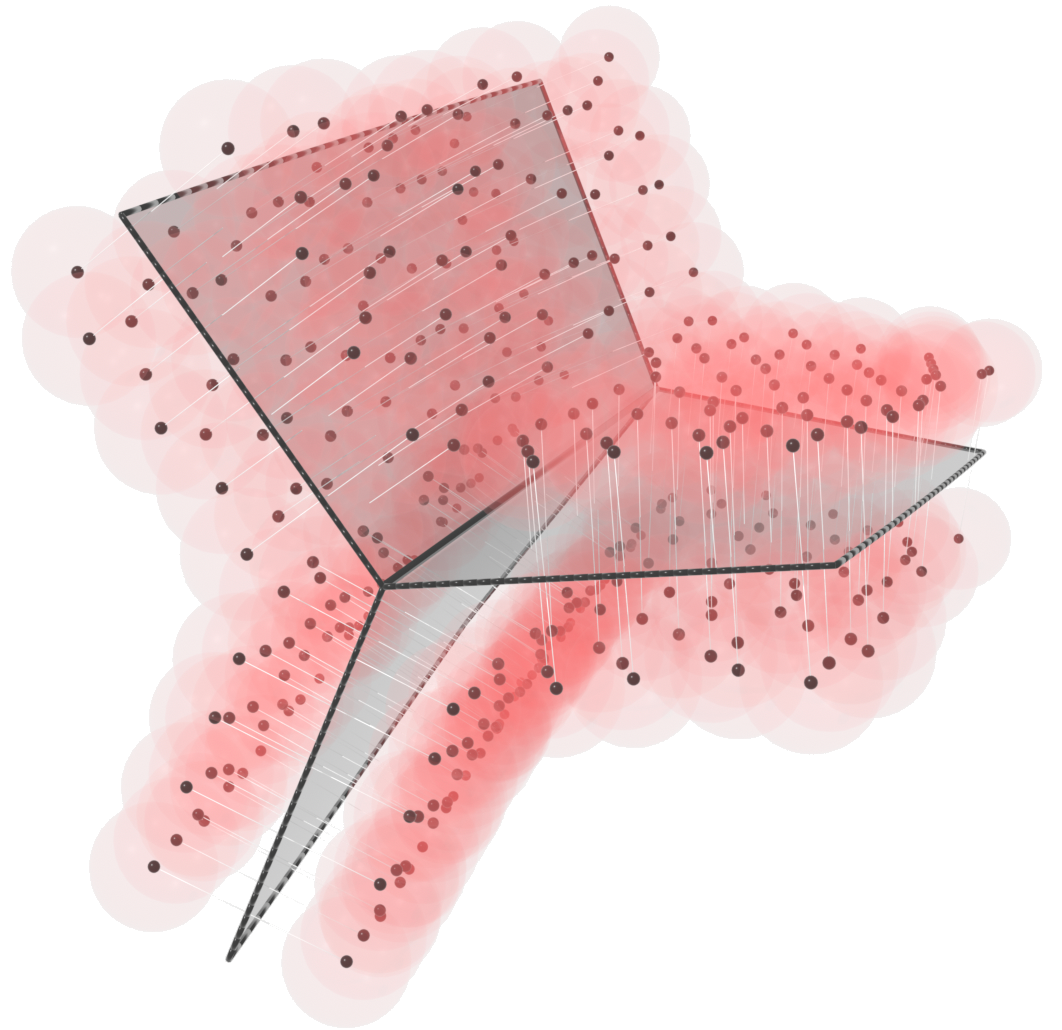}
\end{wrapfigure}
In the first stage of our pipeline, we generate samples on the offset surface $\partial\Omega_\alpha=\{x\in\mathbb{R}^3\mid \phi(x)=\alpha\}$ directly from the queryable UDF.
We adopt the ray-based strategy of \cite{raycastSampling}, who show that area-uniform sampling on an implicit surface can be obtained by casting random rays in space and taking all their intersections with the surface.
They show how this strategy avoids the classical sampling biases of grid-based methods while requiring fewer function evaluations.
Concretely, random rays are cast through an expanded axis-aligned bounding box of the shape and we search for the intersections between each ray and $\partial\Omega_\alpha$ using a sphere-tracing algorithm.
To capture all crossings along the same ray, a small step is taken past the last intersection and the marching continues until the next intersection is found or exiting the box.
These samples constitute an initial set of generators, scattered over potentially different connected components of the shape, from which we refine the sampling using a Poisson disk sampling procedure, resulting in well-spaced samples on $\partial\Omega_\alpha$ (see inset figure).
The Poisson disk radius $r$ should not exceed $\alpha$ which corresponds to the local feature size of $\Omega_\alpha$.

Each obtained sample $\mathbf{x}_i$ is associated with its normal vector $\mathbf{n}_i$ w.r.t. the offset surface, which provides the local tangent information required by our SQEM-based fitting metric in the subsequent stage.
In the ideal case of an exact UDF, the gradient $\nabla\phi(\mathbf{x}_i)$ should provide correct unit normal vectors.
However, in many concrete situations this assumption of exactitude does not hold, notably with neural representations produced by a learning process or with Q-MDF, which are inherently not correct distance fields.
Therefore, we compute these normal vectors by first building a $k$NN graph of the samples (we use $k=10$ throughout all our experiments).
As surface samples from opposite sides of the shape may be close to each other, the graph is filtered so that two samples $\mathbf{x}_i$ and $\mathbf{x}_j$ are connected only if the angle between $\nabla\phi(\mathbf{x}_i)$ and $\nabla\phi(\mathbf{x}_j)$ is below a given threshold.
A plane is then fitted to the connected neighbours of each sample, yielding its associated normal vector.
We denote the resulting set of oriented samples by $\mathcal{V}_\alpha=\{(\mathbf{x}_i,\mathbf{n}_i)\}$.

\subsection{Spheres initialization}
\label{sec:initialization}

In the next step, we efficiently generate an initial set of well-dis\-tri\-bu\-ted medial axis samples, i.e. medial spheres.
First, each sample of $\mathcal{V}_\alpha$ is associated with a candidate medial sphere.
If the UDF is exact, the offset surface $\partial\Omega_\alpha$ would have a constant local feature size of $\alpha$, i.e. each point on the offset surface would lie at distance exactly $\alpha$ from the medial axis.
Under this assumption, a natural way to obtain a medial sphere from a given surface sample $(\mathbf{x}_i,\mathbf{n}_i)$ is to displace it by $\alpha$ along its inward normal direction and define a sphere $m_i=(\mathbf{c}_i,r_i)$, with center $\mathbf{c}_i=\mathbf{x}_i-\alpha\,\mathbf{n}_i$ and radius $r_i=\alpha$.
As stated in Section~\ref{sec:sampling}, this assumption is rarely true on actual data.
The $\alpha$-offset layer is not perfectly uniform, and the local feature size of $\partial\Omega_\alpha$ may deviate from the ideal value $\alpha$.
To compensate for these variations, we instead compute the medial ball for each offset sample using the shrinking ball algorithm \cite{ma2012shrinkingball}.
\setlength{\columnsep}{8pt}
\setlength{\intextsep}{0pt}
\begin{wrapfigure}{r}{0.5\columnwidth}
    \centering
    \includegraphics[width=0.5\columnwidth]{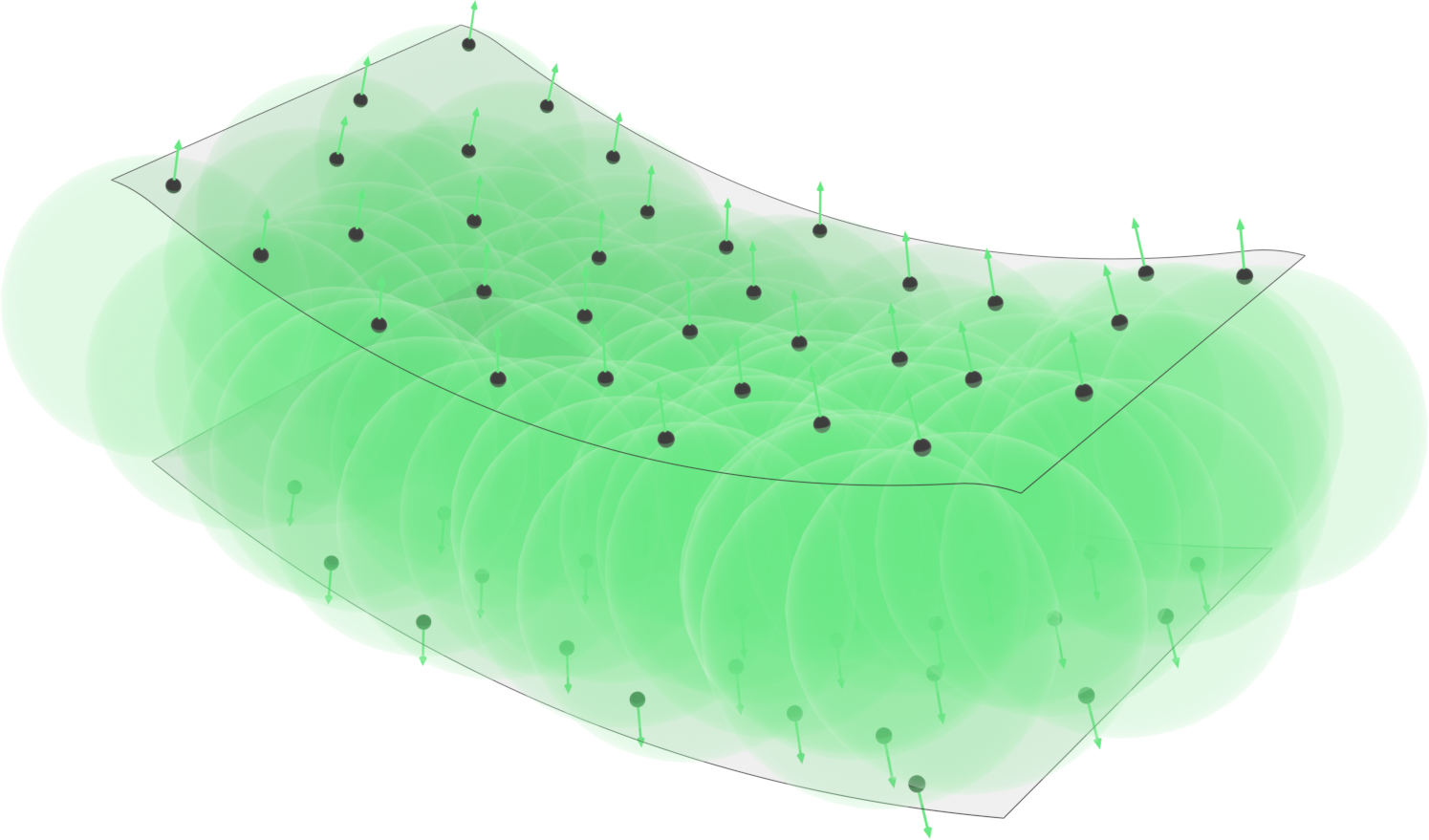}\\
    \includegraphics[width=0.5\columnwidth]{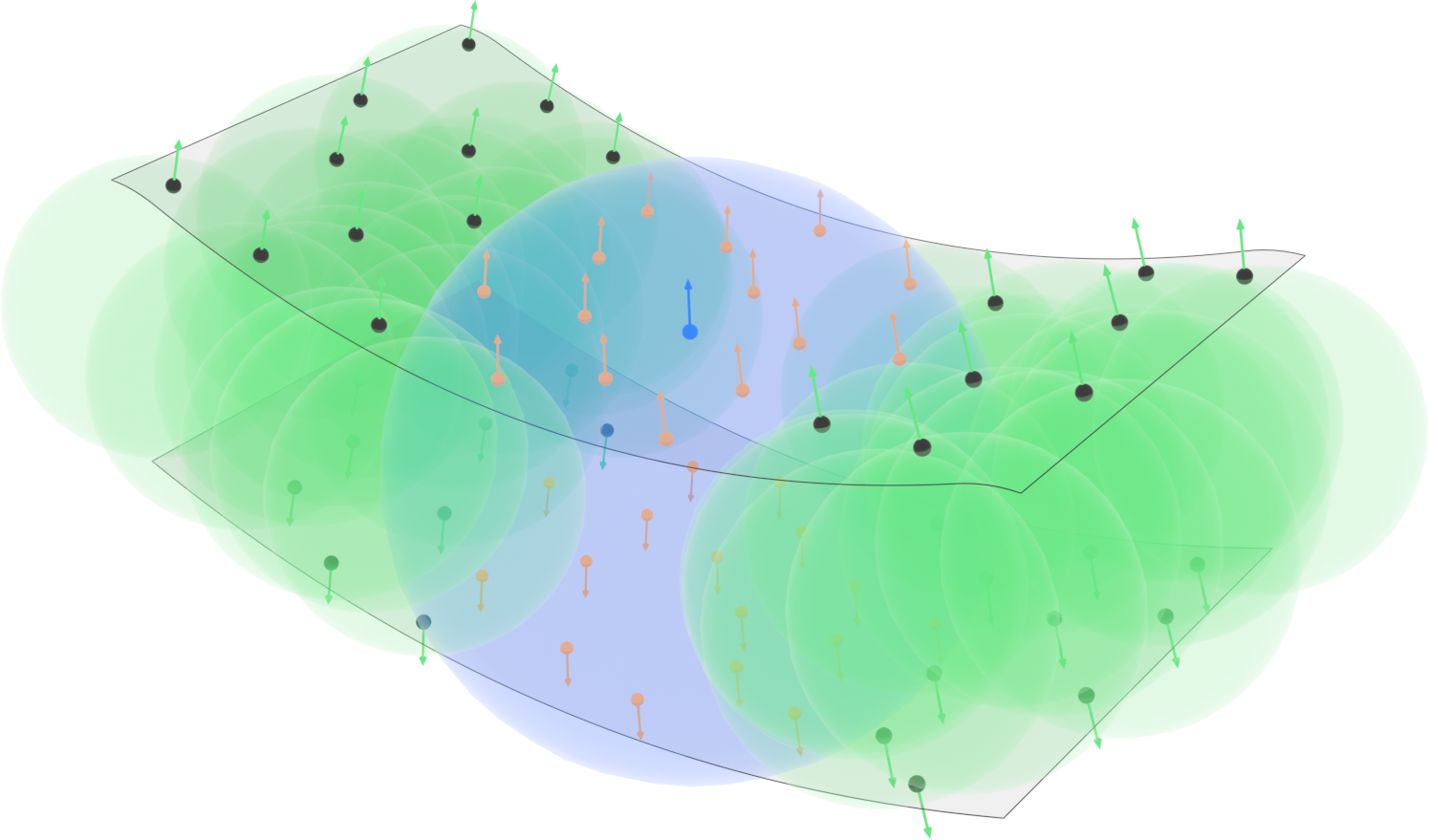}
\end{wrapfigure}

A subset of these candidate spheres is then selected through a simple coverage strategy, inspired by~\cite{dou2022coverage}.
For each sphere $m_i$, we enlarge its radius to $\tilde r_i=r_i+\delta$, where $\delta>0$ is a small constant.
A sample $\mathbf{x}$ is covered by $m_i$ if $\|\mathbf{x}-\mathbf{c}_i\|\le \tilde r_i$.
The goal is to extract a subset of candidate spheres whose dilated supports collectively cover all the samples of $\mathcal{V}_\alpha$.
To this end, the candidate medial spheres are first sorted by decreasing radius.
In an iterative process, the largest remaining sphere is selected and the samples covered by its dilated support are removed from further consideration, together with their associated candidate medial balls (see inset figure).
The process finishes when all samples are covered.
To compute coverage efficiently, we use a flood-filling strategy over the samples' $k$NN graph.
For each selected sphere $m_i$, its two nearest samples in $\mathcal{V}_\alpha$, which are obtained directly during the shrinking ball computation, are used as seeds of a breadth-first traversal on the $k$NN graph.
For each visited sample, its neighboring samples are considered as covered whenever the above condition is satisfied.
Only covered neighbors are pushed into the queue, so the traversal naturally stops at the boundary of the sphere's influence region.
The effect of this parameter on the result is discussed in the experiments section.

\subsection{Spheres optimization}
\label{sec:sphere_fitting_clustering}
This initialization provides a well-distributed set of medial spheres, yet their parameters require refinement to accurately capture the local geometry of the offset volume $\Omega_\alpha$.
Sphere parameters are therefore optimized through a local geometric fitting step inspired by VMAS \cite{VMAS}.
Before the first sphere update, we initialize each cluster $\mathcal{C}_i$ by assigning every offset-surface sample to the sphere with the lowest fitting cost.
An iterative optimization scheme alternately computes a clustering of the offset surface samples $\mathcal{V}_\alpha$, and updates each sphere by fitting it to its current cluster.

\paragraph{Sphere update}
For a sphere $m_i=(\mathbf{c}_i,r_i)$ and its cluster $\mathcal{C}_i$ of oriented samples $v_j=(\mathbf{x}_j,\mathbf{n}_j)$, the fitting step can be formulated as a minimization of the spherical quadric error metric (SQEM) \cite{thiery2013sphere}, which measures the squared distance between the sphere and the tangent planes induced by the samples:
\begin{equation}
E_{\mathrm{SQEM}}(m_i)=\sum_{v_j \in \mathcal{C}_i} Q_{v_j}(c_i, r_i),
\end{equation}
where $Q_{v_j}$ denotes the quadric associated with the tangent plane at sample $v_j$.

However, as acknowledged in recent works \cite{VMAS}, \cite{wang2025matstruct}, this metric does not admit unique optimal spheres in degenerate cases such as locally planar or cylindrical regions where tangent-plane constraints leave degrees of freedom to the sphere center.
VMAS addresses this issue by adding an additional Euclidean regularization term between the sphere and the surface samples.
While this stabilizes the optimization, it also tends to enlarge the sphere radius and turns the objective into a non-linear problem, which must then be solved iteratively using methods such as Gauss--Newton.

We propose instead to incorporate the \emph{line quadric}~\cite{lineQuadric}, originally introduced to resolve QEM degeneracies in surface mesh simplification.
Each sample $v_j=(\mathbf{x}_j,\mathbf{n}_j)$ defines a normal line passing through $\mathbf{x}_j$ with direction $\mathbf{n}_j$.
The corresponding line quadric $L_{v_j}$ measures the squared distance from a point in $\mathbb{R}^3$ to this line.
Using this measure as a constraint acting uniquely on the sphere center, we define the local fitting energy for sphere $m_i$ as:
\begin{equation}
E_i=
\sum_{v_j \in \mathcal{C}_i}
\left(
Q_{v_j}(c_i, r_i)+\mu\,L_{v_j}(\mathbf{c}_i)
\right),
\end{equation}
where $Q_{v_j}$ is the SQEM term, $L_{v_j}$ is the line quadric term, and $\mu$ balances their relative contributions.
In all our experiments, we used a value of $\mu=0.2$.

\setlength{\columnsep}{8pt}
\setlength{\intextsep}{0pt}
\begin{wrapfigure}{r}{0.5\columnwidth}
    \centering
    \includegraphics[width=0.4\columnwidth]{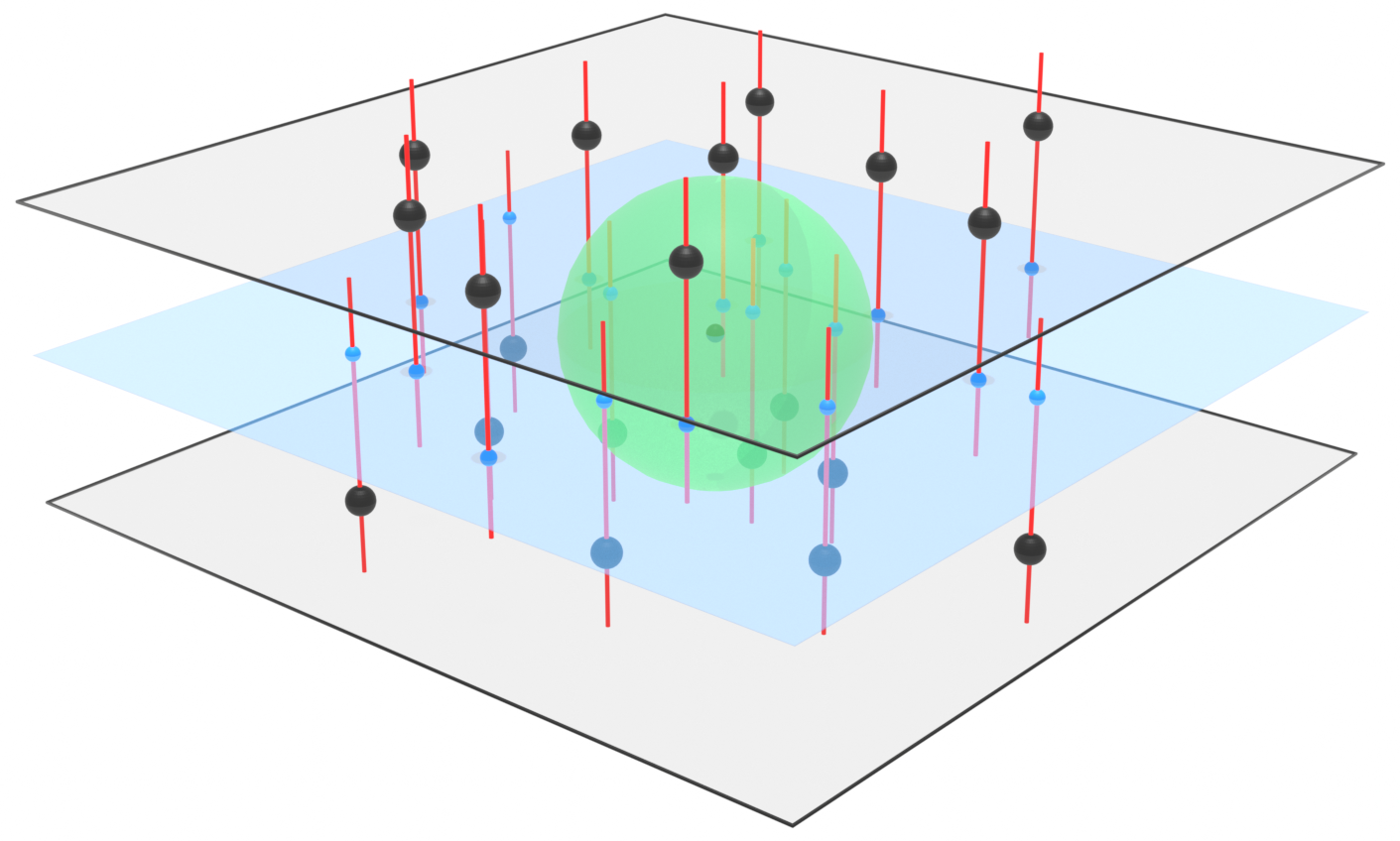}
    \includegraphics[width=0.35\columnwidth]{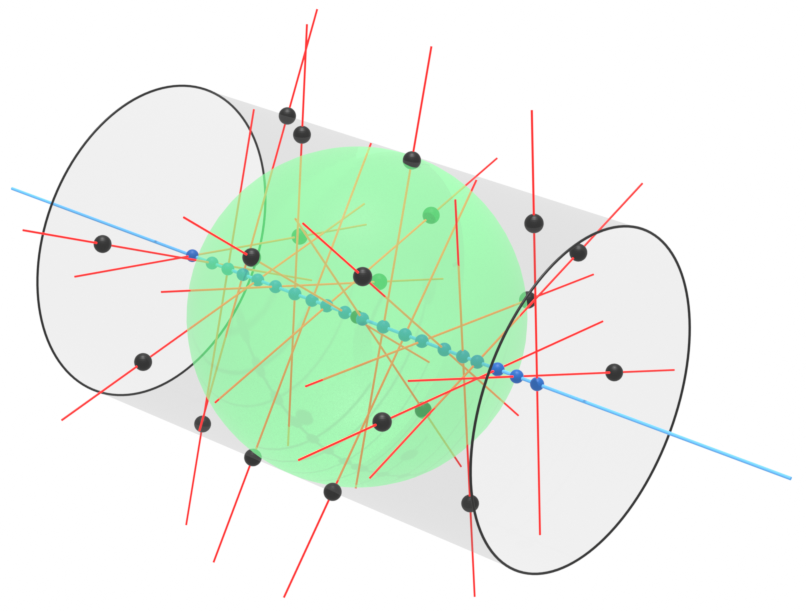}
\end{wrapfigure}
The line quadric provides a stable constraint in locally degenerate configurations (see inset figure).
\emergencystretch=1em
In \textit{planar} regions, the normals are parallel, and the line quadric constraints amount to point-to-point quadratic distances on the medial plane, encouraging sphere centers to distribute evenly across the planar medial region.
In \textit{cylindrical} regions, the normal lines intersect near the center of the circular cross-section of the cylinder.
Consequently, when multiple samples are considered, their combined line-quadric terms encourage the centers to lie along the cylinder's medial axis while remaining well distributed along it.
In the ideal case of an exact UDF, the local thickness of $\Omega_\alpha$ should be constant.
Therefore, the radius of each medial sphere should be equal to $\alpha$, making radius optimization unnecessary.
However, as discussed in previous sections, for robustness to imperfect data, we optimize both the sphere center and its radius:
\begin{equation}
   (\hat{\mathbf{c}}_i,\hat{r}_i)=\arg\min_{\mathbf{c}_i,r_i} E_i
\end{equation}
Since both the SQEM and line quadric terms are quadratic, this update is a small linear least-squares problem and admits a closed-form solution.

Allowing the radius to vary freely, however, may lead in practice to invalid solutions in rare specific cases: the optimized radius can become negative in concave regions, or grow without bound when the cluster is nearly planar (i.e. T1 spheres \cite{wang2022computing}).
To guard against such cases, we define a local upper bound $\bar r_i = 1.5\, r_{n(i)}$, where $n(i)$ denotes the neighbor sphere whose center is closest to $\mathbf{c}_i$.
The update is then accepted only if $0 < \hat r_i \le \bar r_i$.
Otherwise, we keep the previous radius and re-optimize only the center:
\begin{equation}
(\mathbf{c}_i^{\kappa+1},r_i^{\kappa+1})=
\begin{cases}
(\hat{\mathbf{c}}_i,\hat r_i), & \text{if } 0 < \hat r_i \le \bar r_i,\\[4pt]
(\displaystyle \arg\min_{\mathbf{c}_i} E_i,\, r_i^\kappa), & \text{otherwise}
\end{cases}
\end{equation}
In practice, we found this strategy to be more robust than projecting the infeasible radius onto the feasible boundary.
Details about these computations are given in Appendix~\ref{app:sphere_update_derivation}.

\begin{table*}[t]
\centering
\scriptsize
\setlength{\tabcolsep}{1pt}
\renewcommand{\arraystretch}{1.15}

\caption{Quantitative comparison on different datasets and resolutions.
CD and HD are reported in units of $10^{-3}$.
Vertex counts are reported in thousands (K).
Lower is better for Chamfer Distance (CD) and Hausdorff Distance (HD).
Higher is better for Triangle Quality (TQ).
Best and second-best metric results are shown in \textbf{bold} and
\underline{underlined}, respectively.
For our method, we list $(\alpha,r,\delta)$ parameters in the same order as the reported resolutions: DeepFashion uses $(0.002,0.002,0.01)$,
$(0.0015,0.0015,0.0075)$, and $(0.001,0.001,0.005)$;
3DScene and ShapeNetCar both use (0.002, 0.002, 0.01) and (0.0015, 0.0015, 0.0075);}

\label{tab:big_comparison}

\begin{tabular}{ll
cccc|
cccc|
cccc|
cccc|
cccc|
cccc|
cccc|
cccc|
}
\toprule

\multirow{2}{*}{Dataset} & \multirow{2}{*}{Res.}
& \multicolumn{4}{c}{CAP-UDF}
& \multicolumn{4}{c}{DualMeshUDF}
& \multicolumn{4}{c}{MeshUDF}
& \multicolumn{4}{c}{GeoUDF}
& \multicolumn{4}{c}{DCUDF}
& \multicolumn{4}{c}{DCUDF2}
& \multicolumn{4}{c}{MIND}
& \multicolumn{4}{c}{Ours} \\

\cmidrule(lr){3-6}
\cmidrule(lr){7-10}
\cmidrule(lr){11-14}
\cmidrule(lr){15-18}
\cmidrule(lr){19-22}
\cmidrule(lr){23-26}
\cmidrule(lr){27-30}
\cmidrule(lr){31-34}

&
& \#V & CD & HD & TQ
& \#V & CD & HD & TQ
& \#V & CD & HD & TQ
& \#V & CD & HD & TQ
& \#V & CD & HD & TQ
& \#V & CD & HD & TQ
& \#V & CD & HD & TQ
& \#V & CD & HD & TQ \\

\midrule

\multirow{3}{*}{DeepFashion} & 64
& 10.7K & 4.14 & \textbf{39.78} & 0.66
& 10.3K & \underline{2.54} & 43.27 & 0.62
& 9.3K & 6.29 & 56.22 & 0.67
& 6.7K & 3.61 & 47.51 & 0.78
& 8.8K & 6.20 & 68.74 & \underline{0.90}
& 9.7K & 4.90 & 57.89 & 0.88
& -- & -- & -- & --
& 6.1K & \textbf{2.18} & \underline{40.95} & \textbf{0.93} \\

& 128
& 44.6K & 2.68 & \underline{33.15} & 0.66
& 48.4K & \textbf{2.10} & 39.44 & 0.65
& 38.2K & 3.40 & 40.65 & 0.67
& 28.3K & 2.13 & \textbf{32.16} & 0.80
& 65.8K & 2.50 & 34.44 & \underline{0.89}
& 80.7K & 2.71 & 35.12 & 0.88
& 23.0K & 4.16 & 46.75 & 0.84
& 12.6K & \underline{2.11} & 40.60 & \textbf{0.94} \\

& 256
& 190.7K & 2.17 & \underline{32.34} & 0.67
& 226.2K & \underline{2.03} & 38.48 & 0.71
& 238.4K & 3.65 & \textbf{31.57} & 0.75
& 132.5K & \textbf{1.85} & 33.25 & 0.80
& 143.8K & 2.12 & 43.81 & \underline{0.89}
& 225.4K & 2.20 & 41.49 & 0.87
& 94.1K & 3.17 & 41.57 & 0.78
& 27.5K & 2.10 & 37.88 & \textbf{0.94} \\

\midrule

\multirow{2}{*}{ShapeNetCar} & 128
& 64.7K & 3.91 & \textbf{28.88} & 0.73
& 58.3K & \underline{3.62} & 37.01 & 0.66
& 48.3K & 4.33 & \underline{32.92} & 0.74
& 38.4K & 3.68 & 35.68 & 0.82
& 66.2K & 4.15 & 35.78 & \underline{0.88}
& 82.9K & 3.97 & 32.95 & 0.88
& 29.2K & 6.37 & 51.94 & 0.83
& 8.57K
& \textbf{3.16}
& 39.34
& \textbf{0.91} \\

& 256
& 285.0K & 3.45 & \textbf{26.83} & 0.74
& 282.5K & 3.50 & 36.05 & 0.70
& 219.4K & 3.50 & \underline{27.80} & 0.76
& 176.9K & \underline{3.23} & 28.67 & 0.84
& 306.6K & 3.35 & 30.35 & \underline{0.87}
& 448.9K & 3.32 & 29.44 & 0.86
& 128.2K & 5.59 & 45.45 & 0.83
& 16.64K
& \textbf{2.95}
& 36.82
& \textbf{0.92} \\

\midrule

\multirow{2}{*}{3DScene} & 128
& 32.8K & \underline{8.36} & 41.45 & 0.71
& 31.3K & \textbf{8.02} & \textbf{38.07} & 0.66
& 26.2K & 8.85 & 50.28 & 0.73
& 20.0K & 8.43 & \underline{38.48} & 0.83
& 42.2K & 9.45 & 42.02 & 0.85
& 53.5K & 9.42 & 41.99 & \underline{0.86}
& 15.7K & 11.24 & 55.65 & 0.82
& 4.57K
& 8.40
& 48.57
& \textbf{0.93} \\

& 256
& 140.2K & \underline{7.78} & 38.64 & 0.71
& 149.0K & 7.99 & \underline{37.05} & 0.71
& 110.2K & 8.87 & 47.25 & 0.75
& 94.3K & \textbf{7.66} & \textbf{36.92} & 0.84
& 175.5K & 8.79 & 38.53 & \underline{0.86}
& 283.7K & 9.27 & 37.73 & 0.85
& 67.6K & 9.43 & 56.26 & 0.83
& 8.70K
& 8.34
& 43.18
& \textbf{0.93} \\

\bottomrule
\end{tabular}
\end{table*}

\paragraph{Cluster update}
Given the current set of spheres, we update the sample-to-sphere assignment using the same per-sample measure as in the fitting step:
\begin{equation}
d(v_j,m_i)=Q_{v_j}(\mathbf{c}_i,r_i)+\mu\,L_{v_j}(\mathbf{c}_i).
\end{equation}
A full reassignment would compare every sample against every sphere, leading to a complexity of $O(NM)$ for $N$ samples and $M$ spheres.
In our setting, where $N$ can reach several hundred thousand and $M$ tens of thousands, this quickly becomes a computational bottleneck.
To reduce the cost of clustering after the first iteration, we perform a local cluster update procedure.
Concretely, we restrict the optimal sphere search to a local candidate set, leveraging the fact that the isotropic line quadric naturally correlates assignment costs with Euclidean proximity.
To this end, an adjacency graph is built over the spheres as the dual of their clusters connectivity.
This is simply obtained by iterating over the edges of the $k$NN graph of samples and connecting two spheres when adjacent samples are assigned to two distinct spheres.
The search for the new lowest cost sphere of a sample is then restricted to its previously associated sphere and its neighbors, significantly reducing the clustering cost while remaining consistent with the current sphere configuration.

The cluster update and sphere update steps are alternated until the change of error is smaller than a predefined threshold ($1e^{-10}$) or a maximum number of iterations ($150$) is reached.

\setlength{\intextsep}{\savedintextsep}
\setlength{\columnsep}{\savedcolumnsep}
\subsection{Mesh Reconstruction}
\label{sec:mesh_reconstruction}

After the sphere optimization has converged, we construct an initial connectivity directly from the sphere adjacency graph.
Each edge of the adjacency graph gives rise to an edge connecting the corresponding sphere centers.
A triangular face is inserted for every triplet of spheres that are pairwise adjacent.
Although simple and efficient, this purely combinatorial construction generally produces many tetrahedral cells, since each 4-clique of mutually adjacent spheres induces a tetrahedron.
In practice, a large fraction of these tetrahedra are nearly degenerate, as the corresponding sphere centers often lie in configurations close to coplanarity.
Inspired by \cite{wang2025matstruct}, we apply a thinning procedure to the mesh to progressively remove its tetrahedral cells.
At each iteration, the procedure removes an incident simplex pair composed of a $d$-simplex and one of its incident $(d\!-\!1)$-simplices, with $d\in\{3,2\}$, that is, either a tetrahedron-face pair or a face-edge pair.
\begin{figure}[tb]
  \centering
  \includegraphics[width=0.60\columnwidth]{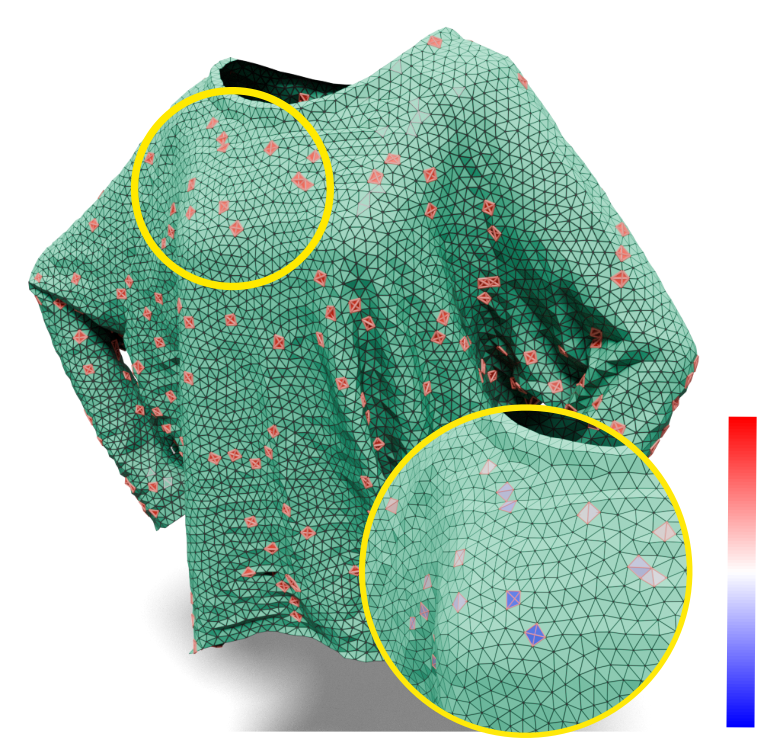}
  \caption{\textbf{Connectivity cleanup:} The initial connectivity extracted from the sphere adjacency graph contains tetrahedral cells, shown in orange. The zoomed-in view colors the faces of a detected tetrahedron according to their integrated UDF scores, which guide simplex-pair removal during thinning.}
  \label{fig:connectivity_cleanup}
\end{figure}
Unlike MATStruct, our deletion order is guided by the integrated UDF value associated with each candidate pair, for both simple and non-simple pairs.
The algorithm alternates between two phases.
In the first phase, all admissible simple pairs are removed in decreasing order of their integrated UDF values, preferentially eliminating simplices located in regions farther from the target surface. When no simple pair remains, we remove the non-simple pair with the largest integrated UDF value.
Such a deletion may expose new simple pairs, in which case the algorithm returns to the simple-pair phase.
This process is repeated until no admissible deletion remains.
Finally, we remove residual isolated non-manifold triangles as a cleanup step.
The UDF score is computed as the area integral of the UDF over the corresponding face.
Figure~\ref{fig:connectivity_cleanup} illustrates this criterion: the detected tetrahedra are shown in orange, and in the zoomed-in view, each face is colored according to its integrated UDF score.

\section{Experiments}
\label{sec:results}

We implemented our method in C++ using the CGoGN library \cite{CGoGN}.
All tests were performed on a computer equipped with an Intel(R) Core(TM) i5-14600 CPU at 3.50 GHz, 32 GB of RAM, and an NVIDIA GeForce RTX 4070 SUPER GPU.
We normalized all input models to the unit cube. Neural UDF models were trained in PyTorch, and the resulting weights were loaded in C++ via LibTorch for subsequent surface reconstruction.

\paragraph{Experimental setup}
We evaluated our method on a diverse set of datasets covering open surfaces, noisy inputs, complex topology, and non-manifold geometry.
For open-surface and noisy inputs, we randomly sampled 100 models from DeepFashion \cite{zhu2020deep}, whose multi-view stereo reconstructions often contain holes, noise, and open boundaries.
For shapes with complex geometry and topology, we randomly selected 100 models from ShapeNet Cars \cite{shapenet2015} and additionally included scenes from the 3D Scene dataset \cite{3Dscene2013}.
To evaluate sharp-feature preservation, we selected 100 CAD models containing sharp features from the ABC dataset \cite{Koch_2019_CVPR}.
For all UDF-based experiments, we used the recent DEUDF \cite{Xu2025DEUDF} to train the underlying UDFs.
To evaluate non-manifold reconstruction, we considered two complementary settings.
First, we randomly sampled 50 models from the ABC dataset along with 10 organic models for which we trained both an SDF and a MF, and extracted the mesh corresponding to the Q-MDF computed as $\mathrm{MF} - |\mathrm{SDF}|$.
Second, following the setup of \cite{ren2025shape}, we constructed 50 synthetic non-manifold models by intersecting pairs of ABC meshes.

\begin{figure}[htb]
    \centering
    \includegraphics[width=0.98\columnwidth]{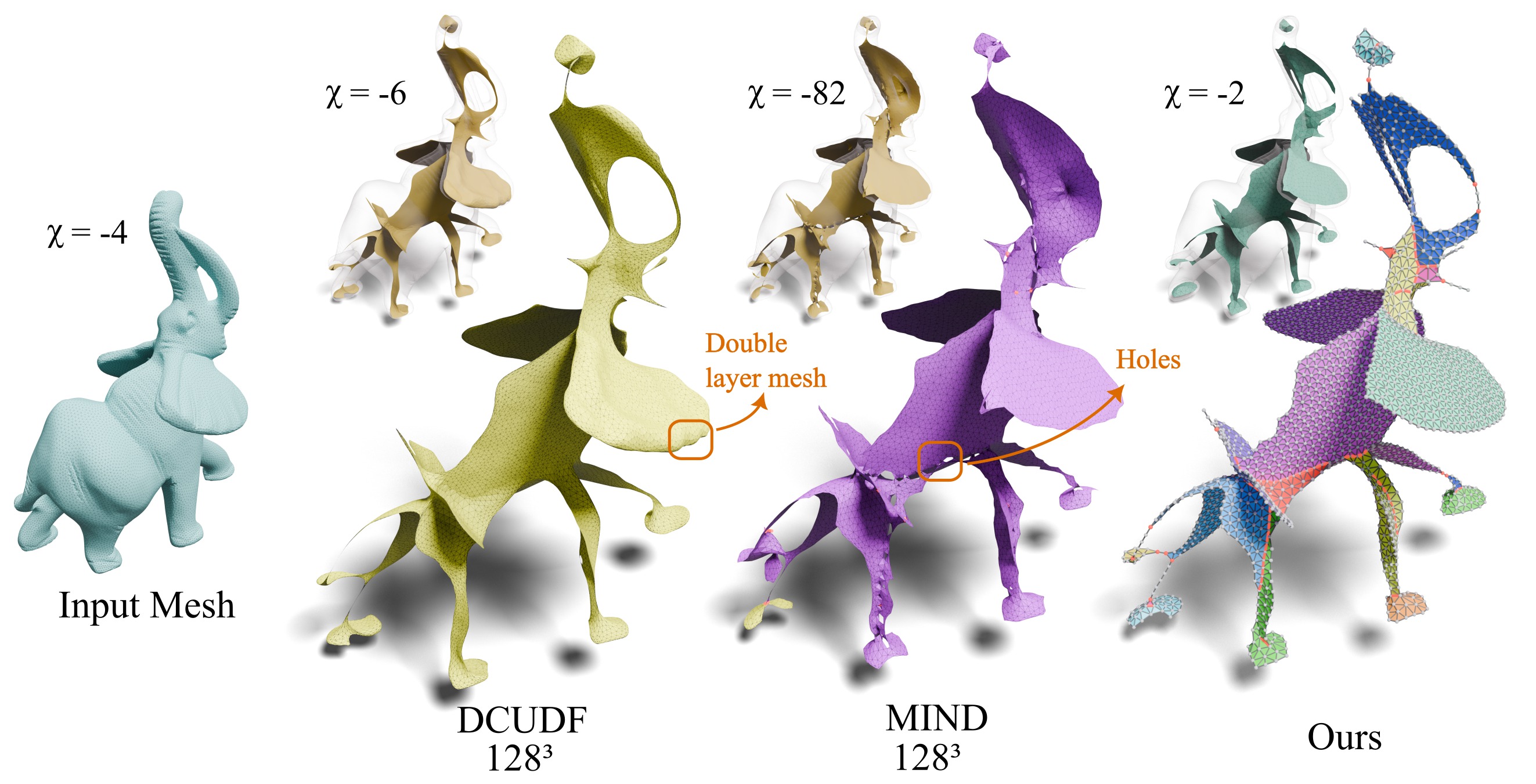}
    \caption{\textbf{Comparison on Q-MDF reconstruction:} For each method, the top row shows the reconstructed mesh, while the bottom row visualizes the corresponding manifold patches in different colors, with non-manifold vertices and edges highlighted in red. The results of DCUDF and MIND are generated at grid resolution $128^3$. Our method uses $\alpha=0.005$, $r=0.002$, and $\delta=0.01$. It successfully recovers the non-manifold edges and partitions the mesh into several coherent manifold patches while DCUDF produces a single manifold patch with duplicated two-layer surfaces and MIND generates visible holes in the reconstruction.}
    \label{fig:elephant}
\end{figure}
\paragraph{Comparisons on UDF reconstruction}
For UDF-based reconstruction, we compare our method against four MC-based extraction methods, CAP-UDF, GeoUDF, MeshUDF and MIND.
Since these methods are originally coupled with their own UDF learning pipe\-lines, we only adopt their mesh extraction algorithms for comparison.
We also compare against one dual-contouring-based method, DualMesh-UDF, and two double-covering methods, DCUDF and DCUDF2.
For UDF-based reconstruction, we evaluate the grid-based baselines at resolutions from $64^3$ to $256^3$, depending on the dataset.
Table~\ref{tab:big_comparison} reports the number of vertices ($\#V$), Chamfer distance (CD), Hausdorff distance (HD), and triangle quality (TQ).
The results show that our method achieves reconstruction accuracy comparable to grid-based baselines while using substantially fewer vertices.
This behavior is expected: in grid-based extraction methods, the geometric detail that can be recovered is directly limited by the grid resolution, whereas our method samples and optimizes the offset surface independently of a fixed voxel grid.
As a result, even with a compact output, we can choose a sufficiently small offset scale to capture fine geometric structures.
In terms of triangle quality, our method consistently produces better results than the baselines, which we attribute to the line-quadric regularization term used during sphere optimization.
\begin{figure}[h]
    \centering
    \includegraphics[width=0.98\columnwidth]{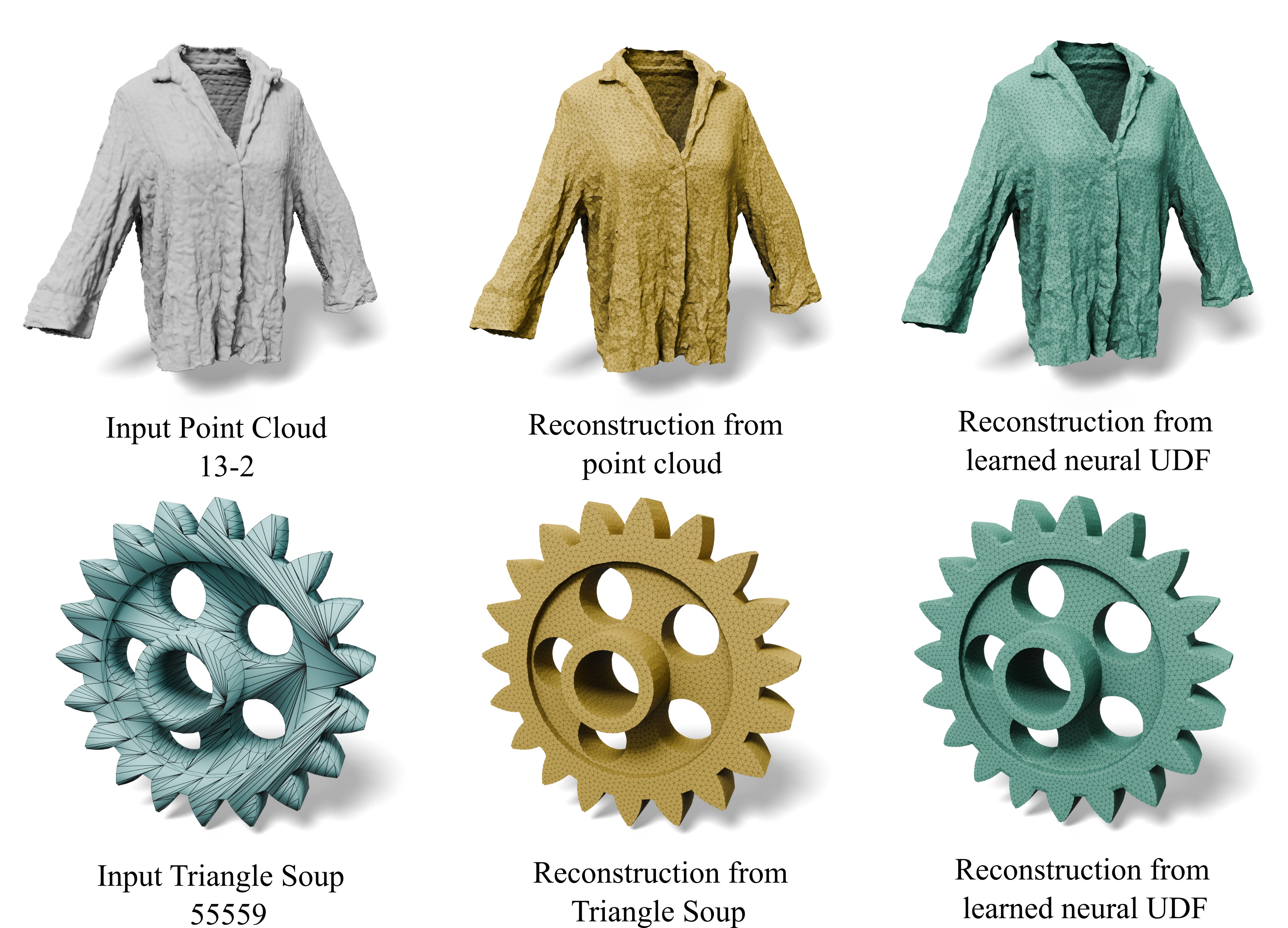}
    \caption{\textbf{Reconstruction from direct inputs.} Both inputs produce results comparable to learned UDFs, while the triangle-soup input better preserves sharp CAD features that are smoothed in the learned representation. Model 55559 is from Thingi10k dataset \cite{Thingi10K}.}
    \label{fig:points_triangles}
\end{figure}
\paragraph{Comparisons on Q-MDF reconstruction}
For Q-MDF reconstruction, we compare our method against DCUDF and MIND.
Since the original implementation of Q-MDF is not publicly available, we use DCUDF as the closest baseline, as both methods follow the same double-covering reconstruction principle.
We further include MIND in the comparison, since it is, to the best of our knowledge, the only existing method that is explicitly designed to reconstruct non-manifold structures from UDFs.
Figure \ref{fig:elephant} shows a qualitative comparison on the elephant model against DCUDF and MIND at grid resolution $128^3$, together with the Euler characteristic of each reconstructed mesh.
To better illustrate the non-manifold structure, we colorize the different manifold patches while highlighting the non-manifold edges in red.
DCUDF produces duplicated two-layer surfaces because its min-cut step cannot be directly applied to non-manifold configurations.
MIND fails to gather sufficient signing information at resolution $128^3$ to reliably assign labels to different regions, resulting in visible holes in the reconstruction and an incorrect Euler characteristic.
In contrast, our method recovers the non-manifold edges, partitions the shape into coherent manifold patches, and preserves the correct topology.
Additional results are provided in the appendix (Figures~\ref{fig:organic} and~\ref{fig:ABC}), where we report DCUDF and MIND results at resolution $256^3$ to illustrate their reconstructions under a higher grid resolution.

\begin{table}[t]
    \centering
    \caption{Sharp-edge reconstruction results on 100 ABC models. All baseline methods use a $128^3$ grid resolution. For our method, we use $\alpha=0.002$, $r=0.002$, and $\delta=0.01$. A dihedral-angle threshold of $45^\circ$ is used to detect sharp feature edges.}
    \label{tab:sharp_edge_cd}
    \small
    \setlength{\tabcolsep}{1pt}
    \renewcommand{\arraystretch}{0.8}
    \begin{tabular}{lcc}
        \toprule
        Method
        & \makecell{Sharp Edge-CD\\($\times 10^{-3}$) $\downarrow$}
        & \makecell{Vertices} \\
        \midrule
        DCUDF       & 74.08 & 43.5K \\
        DCUDF2      & 40.99 & 49.0K \\
        CAPUDF      & 19.42 & 46.0K \\
        GeoUDF      & 35.45 & 36.4K \\
        MeshUDF     & 20.21 & 45.3K \\
        DualMeshUDF & 28.33 & 47.8K \\
        MIND        & 99.53 & 27.8K \\
        \textbf{Ours}
                    & \textbf{12.89}
                    & 22.6K \\
        \bottomrule
    \end{tabular}
\end{table}

\paragraph{Diverse input types}
Our method is not restricted to neural implicit representations and can be applied to any input from which the $\alpha$-offset surface can be sampled.
We demonstrate this flexibility by directly using a point cloud and a triangle soup as input (Figure \ref{fig:points_triangles}) and evaluating the distance function using a Kd-tree and a BVH, respectively.
The reconstructed meshes are comparable to those obtained from UDFs learned from these models, while better preserving sharp CAD features that were smoothed out by the learned UDF representation.

\begin{table}[h]
    \centering
    \caption{Non-manifold edge reconstruction results on 50 intersected
    ABC models. All baseline methods use a $128^3$ grid resolution. For our method, we use $\alpha=0.002$, $r=0.002$, and $\delta=0.005$.}
    \label{tab:nonmanifold_edge_cd}
    \small
    \setlength{\tabcolsep}{1pt}
    \renewcommand{\arraystretch}{0.8}
    \begin{tabular}{lcc}
        \toprule
        Method
        & \makecell{Non-manifold Edge-CD\\($\times 10^{-3}$) $\downarrow$}
        & \makecell{Vertices\\($K$)} \\
        \midrule
        DualMeshUDF
            & 65.64 & 80.2K \\
        MIND
            & 44.18 & 51.5K \\
        Ours(Neural UDFs)
            & 27.69 & 34.7K \\

        \textbf{Ours(Triangle Soup)}
            & \textbf{18.13}
            & 31.5K \\
        \bottomrule
    \end{tabular}
\end{table}
\begin{figure}[htb]
    \centering
    \includegraphics[width=\linewidth]{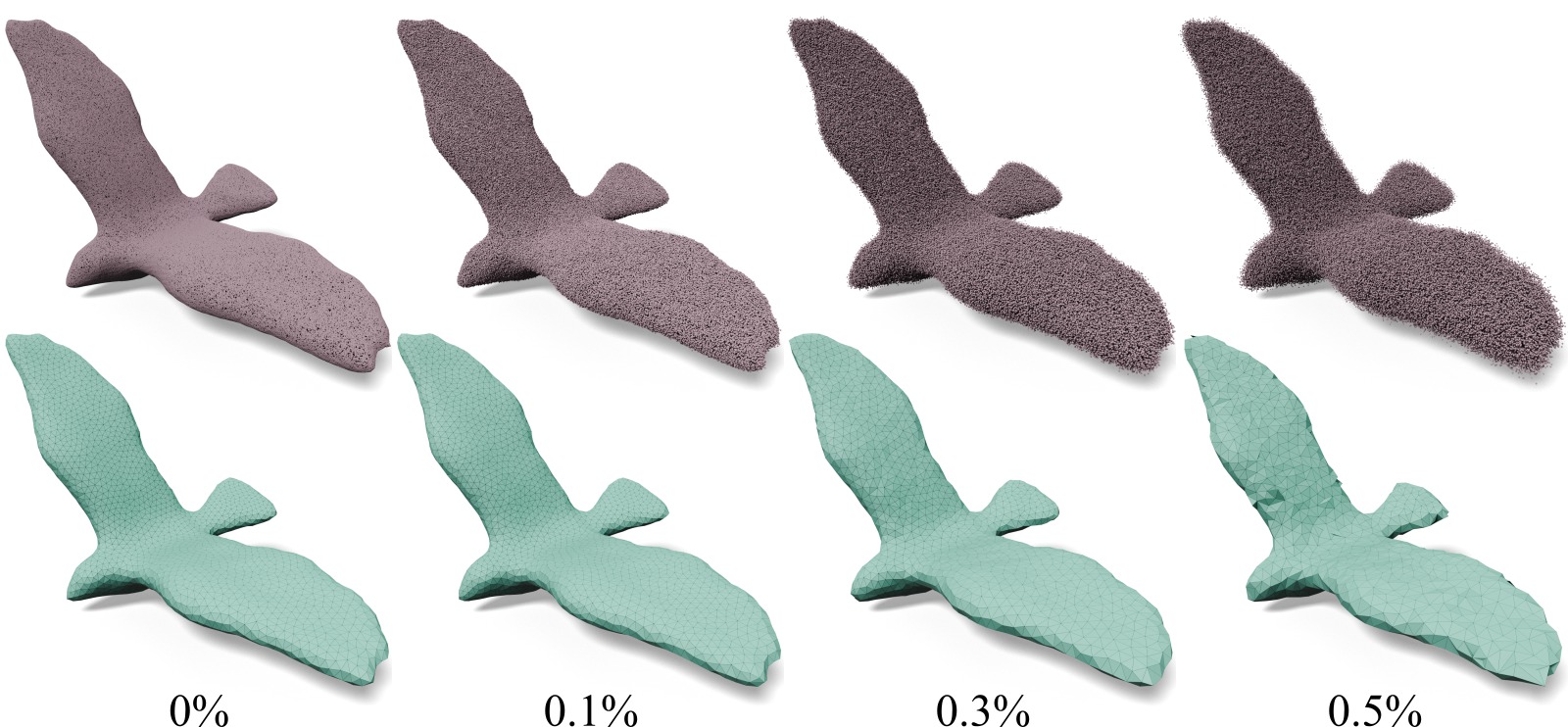}
    \caption{\textbf{Robustness to noisy inputs.} We perturb the input point cloud by random displacements with magnitudes of $0.1$, $0.3$, and $0.5$ of the bounding-box diagonal, and train a UDF from each noisy input. Despite the increasing noise level, our method still produces reasonable reconstructions that preserve  the geometric structures well.}
    \label{fig:noise}
\end{figure}

\paragraph{Feature reconstruction.}
To further assess the preservation of sharp features of CAD model, we report the Edge-CD results in Table \ref{tab:sharp_edge_cd} and provide additional illustrations in Figure~\ref{fig:sharp_feature}.
Our method achieves the best result among all compared approaches while producing a compact reconstruction, demonstrating its ability to accurately localize sharp feature edges from learned UDFs.

In addition to the non-manifold reconstruction quality already shown on Q-MDF reconstructions, we measure the Non Manifold Edge-CD on several synthetic models obtained by intersecting meshes.
These models contain complex and closely spaced intersection curves that are difficult to represent faithfully with neural UDFs.
We thus report results using both learned UDFs and triangle soups as input representations.
Our method achieves the best result among the compared approaches in the learned-UDF setting.
Directly querying the distance to the triangle soups through a BVH avoids neural approximation errors and further improves reconstruction accuracy, yielding the best results both qualitatively and quantitatively.
These results demonstrate the ability of our method to accurately recover both sharp features and complex non-manifold junctions.

\paragraph{Robustness to noise}
We perturb the input point cloud by random displacements with magnitudes of $0.1\%$, $0.3\%$, and $0.5\%$ of the bounding-box diagonal, and train a UDF from each noisy input.
Despite the increasing noise level, our method still produces reasonable reconstructions that preserve the main geometric structures (Figure \ref{fig:noise}).
This stability stems from our cluster-based sphere optimization, where each sphere update aggregates information from multiple local samples and is therefore less sensitive to individual noisy perturbations.
\begin{figure}[htb]
    \centering\includegraphics[width=\linewidth]{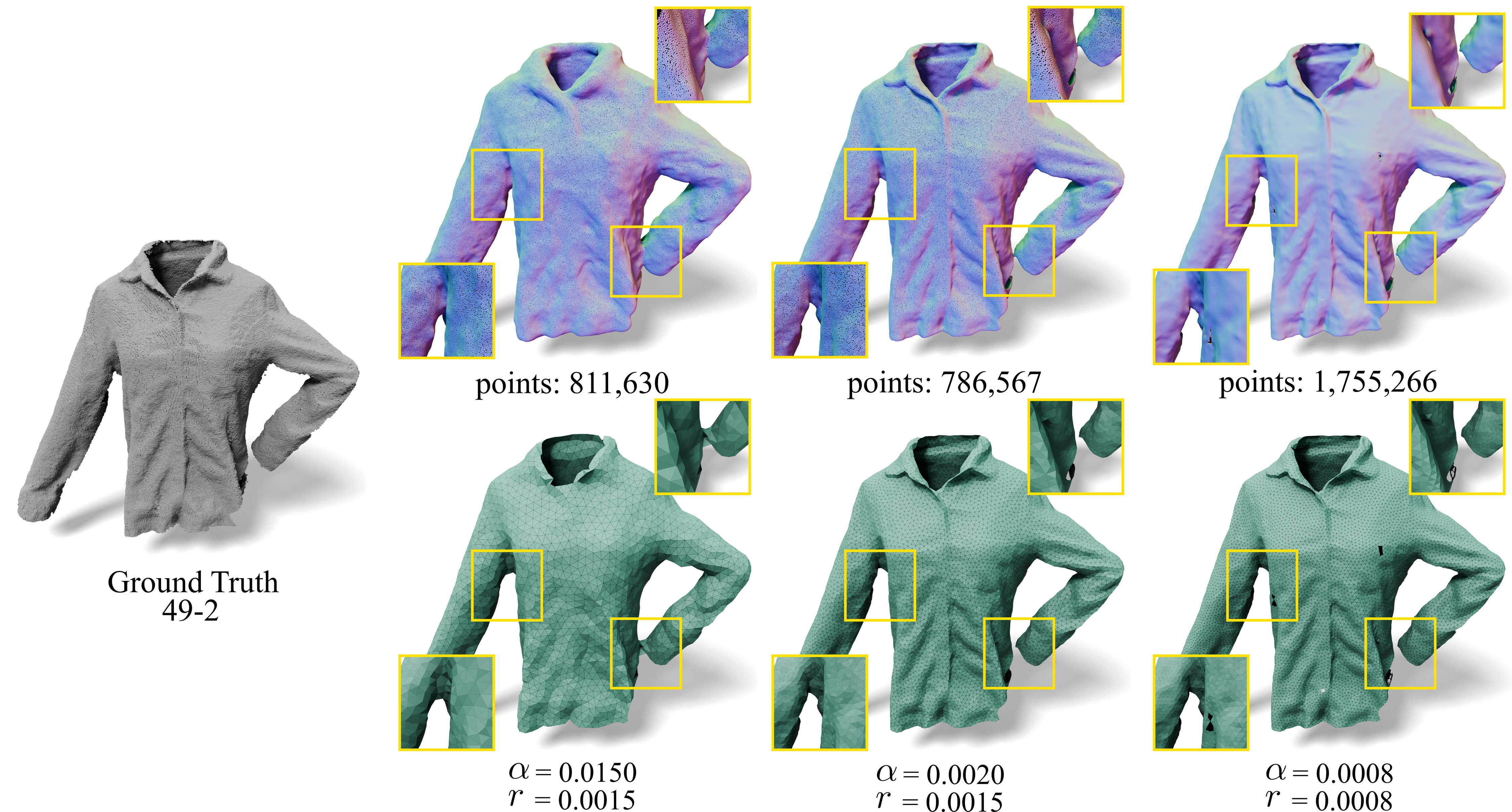}
    \caption{\textbf{Effect of the $\alpha$ parameter:} Top row: sample points on the $\alpha$-level set. Bottom row: the corresponding reconstructions. An excessively large value ($\alpha = 0.015$) can cause the merging of nearby geometric features while an excessively small value ($\alpha = 0.0008$) may introduce holes in the reconstructed mesh if the distance function is not precise enough near the shape. For small $\alpha$, we reduce $r$ to satisfy $r < \alpha$, resulting in denser $\alpha$-level set sampling. The reconstructions are obtained by setting $\delta = 0.0075$.}
    \label{fig:ablation_alpha}
\end{figure}
\begin{figure}[h]
    \centering
    \includegraphics[width=\linewidth]{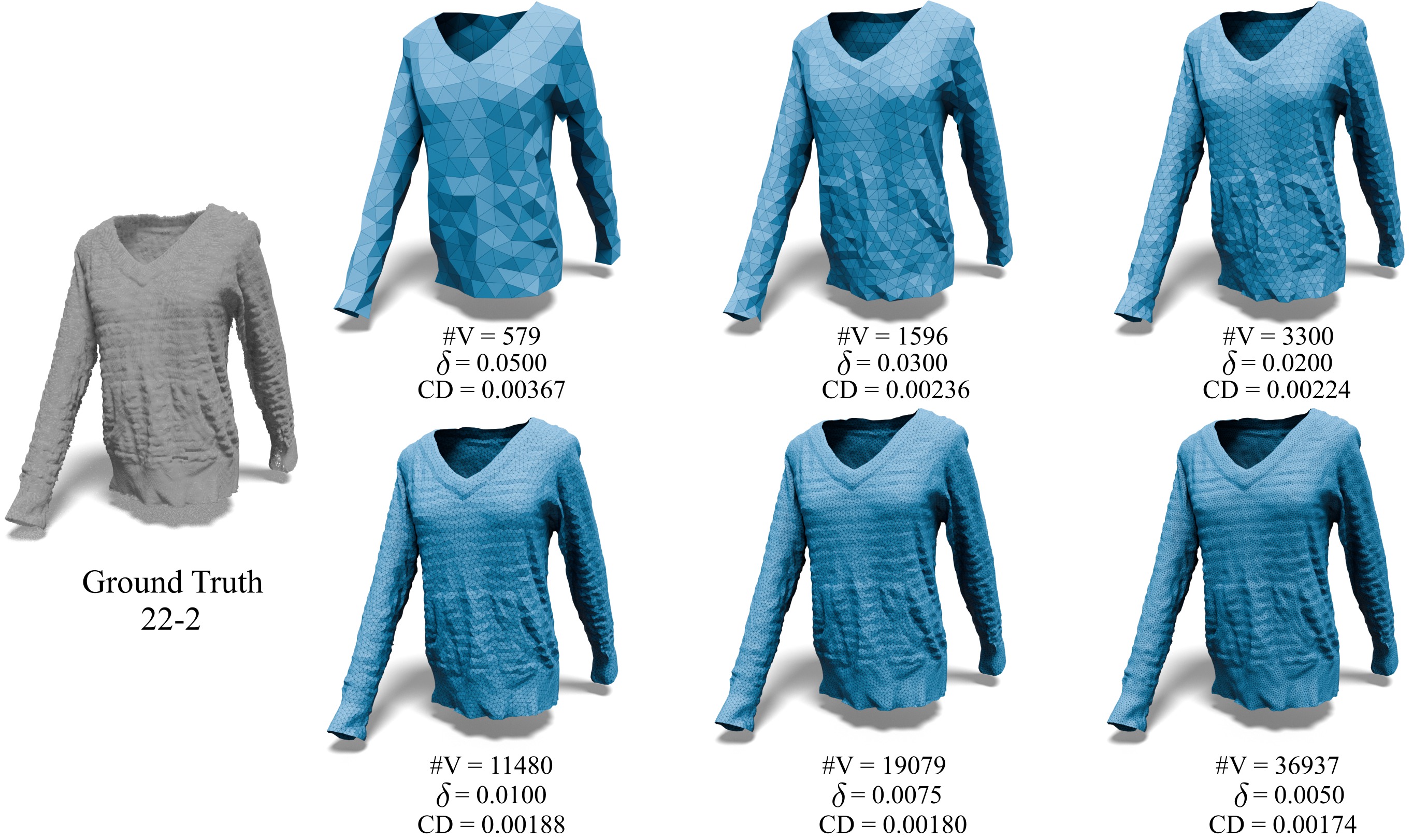}
    \caption{\textbf{Effect of the $\delta$ parameter:} As $\delta$ decreases, the initialization selects a denser set of spheres, which leads to reconstructed meshes with more vertices and captures finer details from the input point cloud. This increased geometric fidelity consistently reduces the Chamfer distance.}
    \label{fig:ablation_delta}
\end{figure}
\paragraph{Parameter study}
The $\alpha$ parameter dictates the distance from the sampled offset surface to the underlying shape.
Selecting an appropriate value is crucial: excessively large values can cause the spurious merging of distinct geometric features, whereas overly small values may introduce holes in the reconstructed mesh (Figure \ref{fig:ablation_alpha}).
This latter issue is particularly visible when processing imperfect distance functions, where the target shape may correspond to non-zero values.
Furthermore, since the sampling radius $r$ is constrained by $\alpha$, decreasing $\alpha$ inherently inflates the number of generated samples, thereby increasing the computational complexity of the pipeline.
As discussed in Section \ref{sec:method}, choosing $\alpha$ requires some prior knowledge of the target shape's scale and characteristics.
The $\delta$ parameter is the dilation constant that controls the number of medial spheres optimized within the offset volume, which directly corresponds to the vertex count of the final extracted mesh.
As demonstrated in Figure \ref{fig:ablation_delta}, smaller values of $\delta$ yield a denser medial axis representation, leading to finer mesh resolution and a correspondingly lower reconstruction error.
We use $\alpha=r=0.002$ and $\delta=0.01$ as a default setting, while these parameters can be adjusted per model for better reconstruction quality.

\begin{figure}[h]
    \centering
    \includegraphics[width=\columnwidth]{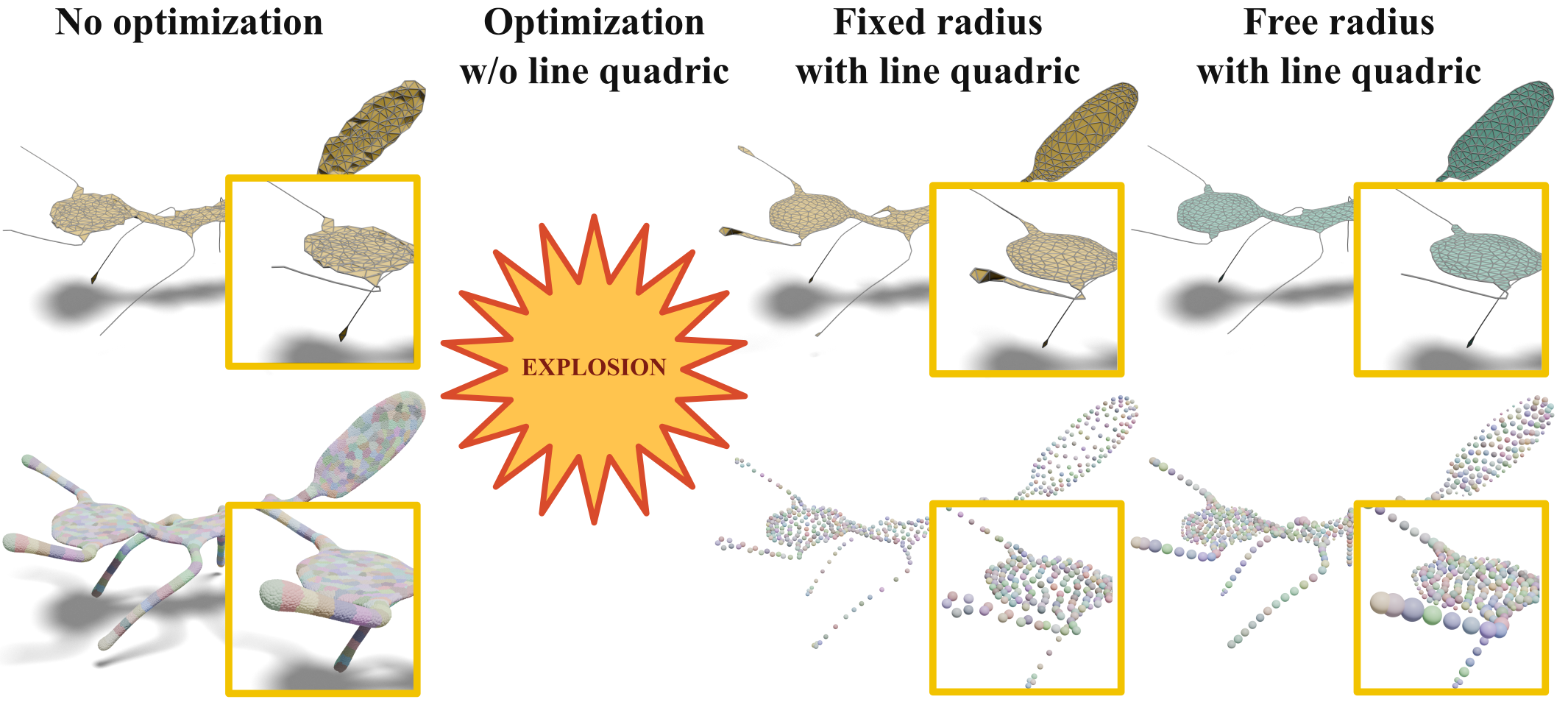}
    \caption{\textbf{Ablation study of optimization}. We compare the result mesh with no optimization, optimization without the line-quadric term, and fixed- versus free-radius optimization.}
    \label{fig:ablation_optimization}
\end{figure}
\paragraph{Ablation study}
We validate the necessity of both the optimization and the free-radius update in Figure \ref{fig:ablation_optimization}.
Without optimization, the resulting mesh contains strongly distorted triangles.
Enabling optimization without regularization term makes the updates unstable and eventually causes the optimization to diverge.
The free-radius update is particularly important when the input UDF is imperfect.
Although we extract the offset at $\alpha=0.005$, the sampled level set does not form a tube of uniform thickness, especially around the bug's antennae.
With fixed radii, the optimizer cannot accommodate this spatially varying geometric scale through radius changes.
Consequently, multiple small spheres accumulate in tubular regions and fail to represent the underlying curve structure cleanly.
Allowing the radii to vary enables the spheres to adapt to the local thickness, producing a more coherent and complete curve skeleton.
It also improves the regularity of the resulting triangulation, as can be clearly observed on the bug's tail.

\begin{figure}[tb]
    \centering
    \includegraphics[width=\columnwidth]{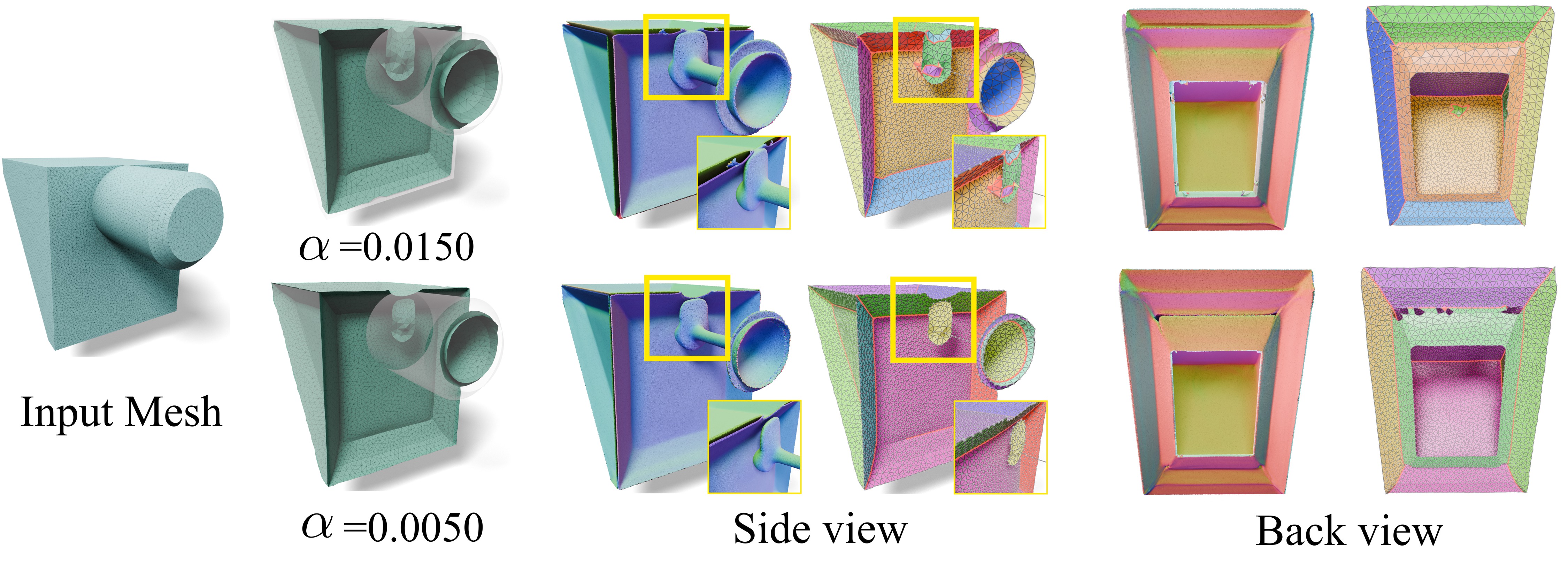}
    \caption{In some challenging situations like this learned Q-MDF of a CAD model, there may not be an ideal $\alpha$ parameter value: an excessive value leads to the merging of some distinct features while a smaller value captures the separated features but introduces holes in some other parts of the shape due to non-zero local minima of the distance function.}
    \label{fig:alpha_limitation}
\end{figure}
\section{Limitations and future work}
\label{sec:limitations}
While our method is designed to better recover open, non-manifold and mixed dimension structures, it does not provide a formal topological guarantee.
This is mainly due to the heuristic nature of our connectivity construction, which is derived from the adjacency of optimized sphere clusters, and to the imperfection of the input data itself.
This issue becomes particularly visible in challenging configurations where no single value of $\alpha$ can produce a defect-free reconstruction.
Figure \ref{fig:alpha_limitation} illustrates this limitation with the example of a Q-MDF characterized by closely spaced geometric structures.
Due to approximation errors, learned Q-MDFs frequently contain regions where the local minimum of the function remains strictly positive.
A sufficiently large $\alpha$ helps our method overcome such positive local minima, but the smaller $\alpha$ needed to separate nearby features inevitably creates holes elsewhere in the reconstruction.
Finally, our current pipeline relies on a globally defined sampling radius and uniform point generation, which produces a homogeneous vertex distribution regardless of local geometric complexity.
A natural direction for future work is to extend the method toward adaptive or anisotropic meshing, reducing the number of elements in low-curvature regions while preserving high resolution near sharp features and complex structures.

\begin{acks}
This research was funded by the Grand Est Region through grant 22-DOC-067 (Def-RV project) and the French government's National Research Agency as part of the Investments for the Future Program (PIA), under grant ANR-22-CE38-0010-01 (POSTURES project). We also thank Dominique Attali for her valuable suggestions and constructive discussions.
\end{acks}

\begin{figure*}[tpb]
    \centering
    \includegraphics[width=0.95\linewidth]{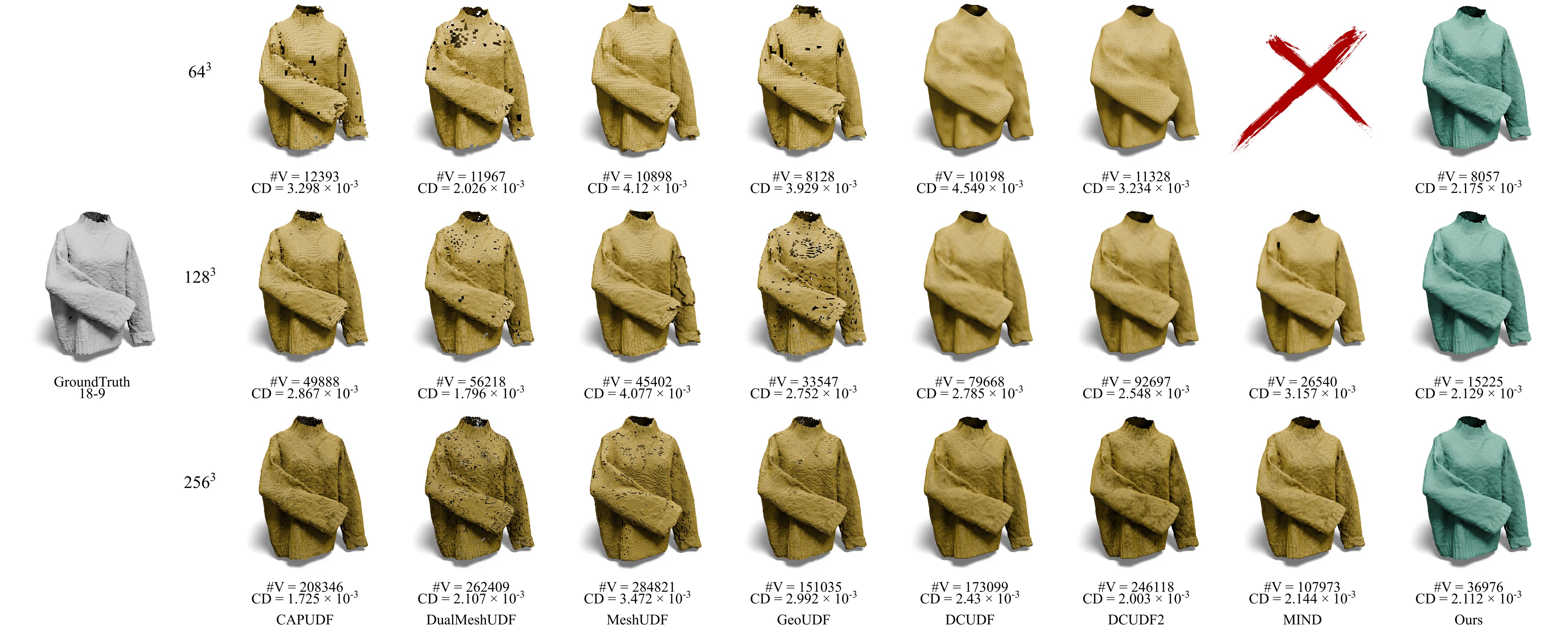}
    \caption{\textbf{Comparison on the DeepFashion dataset.}
We compare our method against seven state-of-the-art methods under three grid resolutions, $64^3$, $128^3$, and $256^3$. For each reconstruction, we report the number of vertices (\#V) and the Chamfer distance (CD).}
    \label{fig:18-9}
\end{figure*}

\begin{figure*}[tpb]
    \centering
    \includegraphics[width=0.95\linewidth]{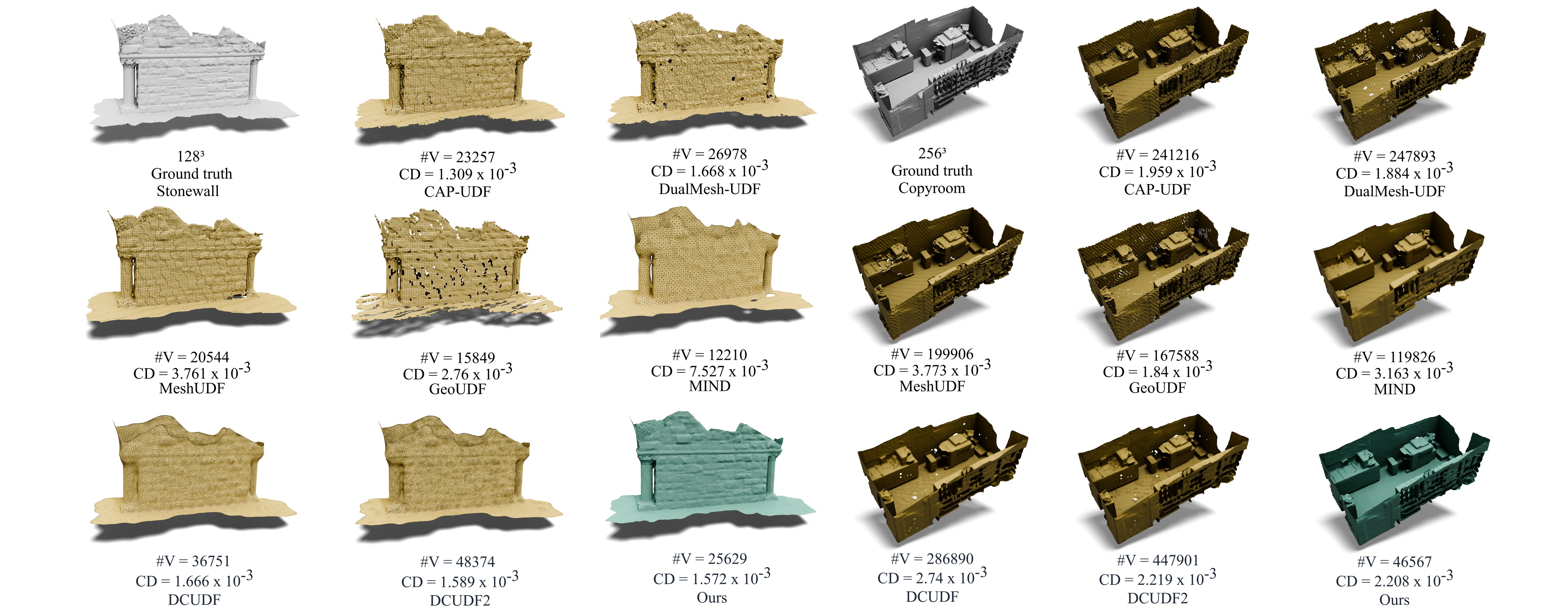}
    \caption{\textbf{Comparison on the 3DScene dataset.}
We compare our method against seven state-of-the-art methods on two models, Stonewall and Copyroom, under two grid resolutions, $128^3$ and $256^3$. For each reconstruction, we report the number of vertices (\#V) and the Chamfer distance (CD).}
    \label{fig:3dScene}
\end{figure*}

\begin{figure*}[tpb]
    \centering
    \includegraphics[width=\linewidth]{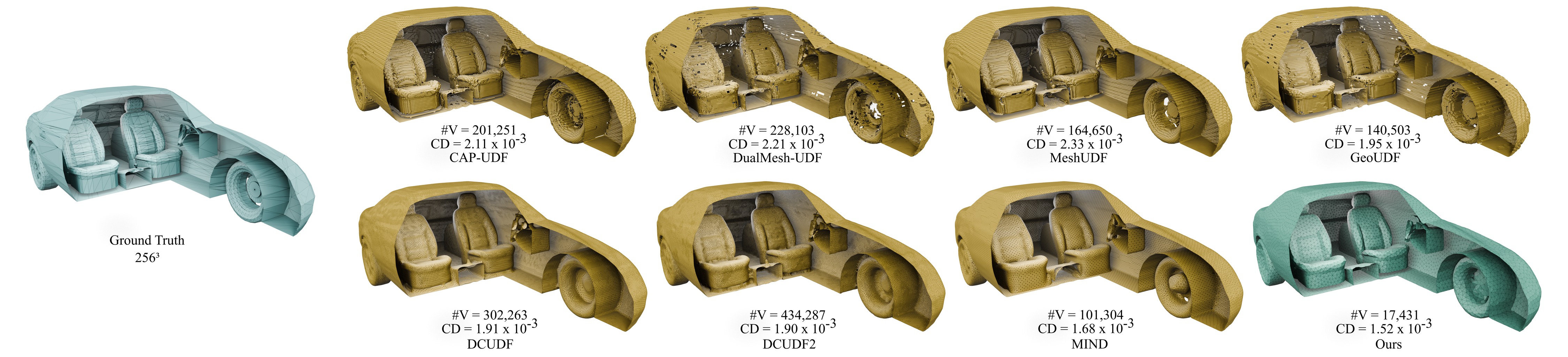}
    \caption{\textbf{Comparison on the ShapeNet Car dataset}. We compare our method against seven state-of-the-art methods under three grid resolutions under the resolution $256^3$. For each reconstruction, we report the number of vertices (\#V) and the Chamfer distance (CD).}
    \label{fig:shapenetcar}
\end{figure*}

\clearpage

\bibliographystyle{ACM-Reference-Format}
\bibliography{offsetaxis}

\clearpage
\appendix

\section{Derivation of the local sphere update}
\label{app:sphere_update_derivation}
\begin{table*}[t]
\centering
\caption{Average mesh extraction time per model (in seconds) for the experiments reported in Table~\ref{tab:big_comparison}.}
\label{tab:extraction_time}
\resizebox{\textwidth}{!}{%
\begin{tabular}{llcccccccc}
\toprule
Dataset & Res.
& CAP-UDF
& DualMeshUDF
& MeshUDF
& GeoUDF
& DCUDF
& DCUDF2
& MIND
& OffsetAxis \\
\midrule

\multirow{3}{*}{DeepFashion}
& 64
& 2.97
& 2.00
& 2.17
& 3.42
& 10.99
& 7.78
& --
& 62.13 \\

& 128
& 10.27
& 2.81
& 4.10
& 9.80
& 45.82
& 50.56
& 57.82
& 158.64 \\

& 256
& 58.95
& 7.72
& 20.77
& 49.49
& 193.84
& 109.99
& 76.53
& 547.15 \\

\midrule

\multirow{2}{*}{ShapeNetCar}
& 128
& 9.06
& 1.96
& 3.13
& 9.06
& 38.61
& 50.36
& 42.11
& 74.36 \\

& 256
& 60.64
& 7.55
& 18.09
& 51.68
& 182.19
& 178.82
& 240.07
& 197.84 \\

\midrule

\multirow{2}{*}{3DScene}
& 128
& 16.88
& 7.32
& 8.69
& 12.77
& 32.65
& 35.59
& 60.04
& 45.08 \\

& 256
& 59.62
& 10.01
& 21.74
& 46.39
& 117.92
& 72.98
& 95.21
& 80.08 \\

\bottomrule
\end{tabular}%
}
\end{table*}

We derive the local update of a sphere $m_i=(\mathbf{c}_i,r_i)$ with respect to a cluster $\mathcal{C}_i$ of oriented samples $v_j=(\mathbf{x}_j,\mathbf{n}_j)$.
Each sample is assigned an area weight $\omega_j=\mathcal{A}(v_j)$.

\subsection{Spherical Quadric Error Metric}
The first term of our metric penalizes the distance from the sphere to the tangent planes of the oriented samples of the cluster using the Spherical Quadric Error Metric~\cite{thiery2013sphere}.
For a given sample $v_j$, the distance from the surface of the sphere $m_i$ to its tangent plane is given by:
\begin{equation}
    d^Q_{v_j}(m_i)
    =
    \mathbf{n}_j^\top(\mathbf{x}_j-\mathbf{c}_i)-r_i
\end{equation}

Representing the sphere by the vector $\mathbf{s}_i = [\mathbf{c}_i^\top, r_i]^\top \in \mathbb{R}^4$, and defining augmented 4D vectors for the sample normal and position as $\tilde{\mathbf{n}_j} = [\mathbf{n}_j^\top, 1]^\top \in\mathbb{R}^4$ and $\tilde{\mathbf{x}_j} = [\mathbf{x}_j^\top, 0]^\top \in\mathbb{R}^4$, the squared distance can be rewritten as:
\begin{equation}
\begin{split}
    d^Q_{v_j}(\mathbf{s}_i)^2
    & =
    (\tilde{\mathbf{n}}_j^\top \tilde{\mathbf{x}}_j - \tilde{\mathbf{n}}_j^\top \mathbf{s}_i)^2\\
    & =
    \mathbf{s}_i^\top (\tilde{\mathbf{n}}_j \tilde{\mathbf{n}}_j^\top) \mathbf{s}_i - 2 (\tilde{\mathbf{n}}_j^\top \tilde{\mathbf{x}}_j \tilde{\mathbf{n}}_j)^\top \mathbf{s}_i + (\tilde{\mathbf{n}}_j^\top \tilde{\mathbf{x}}_j)^2\\
    & =
    \frac{1}{2}\mathbf{s}_i^\top \mathbf{A}^Q_{v_j} \mathbf{s}_i - (\mathbf{b}^Q_{v_j})^\top \mathbf{s}_i + c^Q_{v_j}
\end{split}
\end{equation}
where:
\begin{equation}
    \mathbf{A}^Q_{v_j} = 2 \tilde{\mathbf{n}}_j \tilde{\mathbf{n}}_j^\top,
    \quad
    \mathbf{b}^Q_{v_j} = 2 (\tilde{\mathbf{n}}_j^\top \tilde{\mathbf{x}}_j) \tilde{\mathbf{n}}_j,
    \quad
    c^Q_{v_j} = (\tilde{\mathbf{n}}_j^\top \tilde{\mathbf{x}}_j)^2
\end{equation}

The total SQEM energy for the cluster $\mathcal{C}_i$ is the weighted sum of these individual sample quadrics:
\begin{equation}
    E^Q_i = \frac{1}{2}\mathbf{s}_i^\top \mathbf{A}^Q_i \mathbf{s}_i - (\mathbf{b}^Q_i)^\top \mathbf{s}_i + c^Q_i
\end{equation}
where:
\begin{equation}
    \mathbf{A}^Q_i = \sum_{v_j \in \mathcal{C}_i} \omega_j \mathbf{A}^Q_{v_j},
    \quad
    \mathbf{b}^Q_i = \sum_{v_j \in \mathcal{C}_i} \omega_j \mathbf{b}^Q_{v_j},
    \quad
    c^Q_i = \sum_{v_j \in \mathcal{C}_i} \omega_j c^Q_{v_j}
\end{equation}

\subsection{Line quadric}
The second term of our metric penalizes the distance from the optimized sphere center to the normal lines passing through the samples position and aligned along their normal vector using line quadrics~\cite{lineQuadric}.
It can be defined by summing two plane quadrics induced by two orthonormal vectors spanning the tangent plane orthogonal to the normal vector.
A simpler expression arises from the fact that for a given sample $v_j$, the distance from the sphere center $\mathbf{c}_i$ to this line is the magnitude of the vector $(\mathbf{c}_i - \mathbf{x}_j)$ projected onto the plane orthogonal to $\mathbf{n}_j$.

Let $\mathbf{P}_j = \mathbf{I} - \mathbf{n}_j\mathbf{n}_j^\top$ be the $3 \times 3$ orthogonal projection matrix for this plane.
To integrate this into our 4D optimization space, we pad $\mathbf{P}_j$ with zeros to form a $4 \times 4$ matrix $\tilde{\mathbf{P}}_j$, effectively ignoring the sphere's radius $r_i$ during the distance computation.

Since $\tilde{\mathbf{P}}_j^\top\tilde{\mathbf{P}}_j=\tilde{\mathbf{P}}_j$, the squared distance from the sphere center to the sample normal line can be written as:

\begin{equation}
\begin{split}
    d^L_{v_j}(\mathbf{s}_i)^2
    & =
    (\mathbf{s}_i - \tilde{\mathbf{x}}_j)^\top \tilde{\mathbf{P}}_j (\mathbf{s}_i - \tilde{\mathbf{x}}_j)\\
    & =
    \mathbf{s}_i^\top \tilde{\mathbf{P}}_j \mathbf{s}_i - 2 (\tilde{\mathbf{P}}_j \tilde{\mathbf{x}}_j)^\top \mathbf{s}_i + \tilde{\mathbf{x}}_j^\top \tilde{\mathbf{P}}_j \tilde{\mathbf{x}}_j\\
    & =
    \frac{1}{2} \mathbf{s}_i^\top \mathbf{A}^L_{v_j} \mathbf{s}_i - (\mathbf{b}^L_{v_j})^\top \mathbf{s}_i + c^L_{v_j}
\end{split}
\end{equation}
where:
\begin{equation}
    \mathbf{A}^L_{v_j} = 2 \tilde{\mathbf{P}}_j,
    \quad
    \mathbf{b}^L_{v_j} = 2 \tilde{\mathbf{P}}_j \tilde{\mathbf{x}}_j,
    \quad
    c_{v_j}^L = \tilde{\mathbf{x}}_j^\top \tilde{\mathbf{P}}_j \tilde{\mathbf{x}}_j
\end{equation}

The total line quadric energy for the cluster $\mathcal{C}_i$ is the weighted sum of these individual sample quadrics:
\begin{equation}
    E^L_i = \frac{1}{2}\mathbf{s}_i^\top \mathbf{A}^L_i \mathbf{s}_i - (\mathbf{b}^L_i)^\top \mathbf{s}_i + c^L_i
\end{equation}
where:
\begin{equation}
    \mathbf{A}^L_i = \sum_{v_j \in \mathcal{C}_i} \omega_j \mathbf{A}^L_{v_j},
    \quad
    \mathbf{b}^L_i = \sum_{v_j \in \mathcal{C}_i} \omega_j \mathbf{b}^L_{v_j},
    \quad
    c^L_i = \sum_{v_j \in \mathcal{C}_i} \omega_j c^L_{v_j}
\end{equation}

\subsection{Free-radius update}
The local fitting energy for the sphere $m_i$ is:

\begin{equation}
\begin{split}
    E_i & = E^Q_i + \mu E^L_i\\
        & = \frac{1}{2} \mathbf{s}_i^T \mathbf{A}_i \mathbf{s}_i - \mathbf{b}_i^T \mathbf{s}_i + c_i
\end{split}
\end{equation}
where:
\begin{equation}
    \mathbf{A}_i = \mathbf{A}^Q_i + \mu\mathbf{A}^L_i,
    \quad
    \mathbf{b}_i = \mathbf{b}^Q_i + \mu\mathbf{b}^L_i
    \quad
    c_i = c^Q_i + \mu c^L_i
\end{equation}

Setting $\nabla_{\mathbf{s}_i} E_i = 0$
gives $\mathbf{A}_i\hat{\mathbf{s}}_i = \mathbf{b}_i$,
and thus the free-radius sphere update is obtained as:

\begin{equation}
    \hat{\mathbf{s}}_i
    =
    \begin{bmatrix}
        \hat{\mathbf{c}}_i\\
        \hat r_i
    \end{bmatrix}
    =
    \mathbf{A}_i^{-1}\mathbf{b}_i
\end{equation}

\subsection{Fixed-radius update}
The free-radius solution is accepted only if $0 < \hat r_i \le \bar r_i$.
Otherwise, we keep the radius from the previous optimization step $r_i^\kappa$ and only update the center.
Using the same accumulated system, we write:

\begin{equation}
    \mathbf{A}_i
    =
    \begin{bmatrix}
        \mathbf{A}_{cc} & \mathbf{a}_{cr}\\
        \mathbf{a}_{cr}^\top & a_{rr}
    \end{bmatrix},
    \qquad
    \mathbf{b}_i
    =
    \begin{bmatrix}
        \mathbf{b}_c\\
        b_r
    \end{bmatrix}
\end{equation}

For a given fixed radius $r_i^\kappa$, the first block row gives:

\begin{equation}
    \mathbf{A}_{cc}\mathbf{c}^{\mathrm{fix}}_i
    +
    \mathbf{a}_{cr} r_i^\kappa
    =
    \mathbf{b}_c
\end{equation}

Therefore we can solve for $\mathbf{c}^{\mathrm{fix}}_i$ from the expression:

\begin{equation}
    \mathbf{A}_{cc}\mathbf{c}^{\mathrm{fix}}_i
    =
    \mathbf{b}_c
    -
    \mathbf{a}_{cr} r_i^\kappa
\end{equation}

The final update of the sphere $m_i$ for the $\kappa+1$ optimization step is:

\begin{equation}
    (\mathbf{c}_i^{\kappa+1},r_i^{\kappa+1})
    =
    \begin{cases}
    (\hat{\mathbf{c}}_i,\hat r_i),
    & \text{if } 0 < \hat r_i \le \bar r_i,\\[4pt]
    (\mathbf{c}^{\mathrm{fix}}_i,r_i^\kappa),
    & \text{otherwise}
    \end{cases}
\end{equation}

\section{Runtime Analysis}
\label{sec:additional_runtime_analysis}

\begin{figure}[t]
    \centering
    \includegraphics[width=\columnwidth]{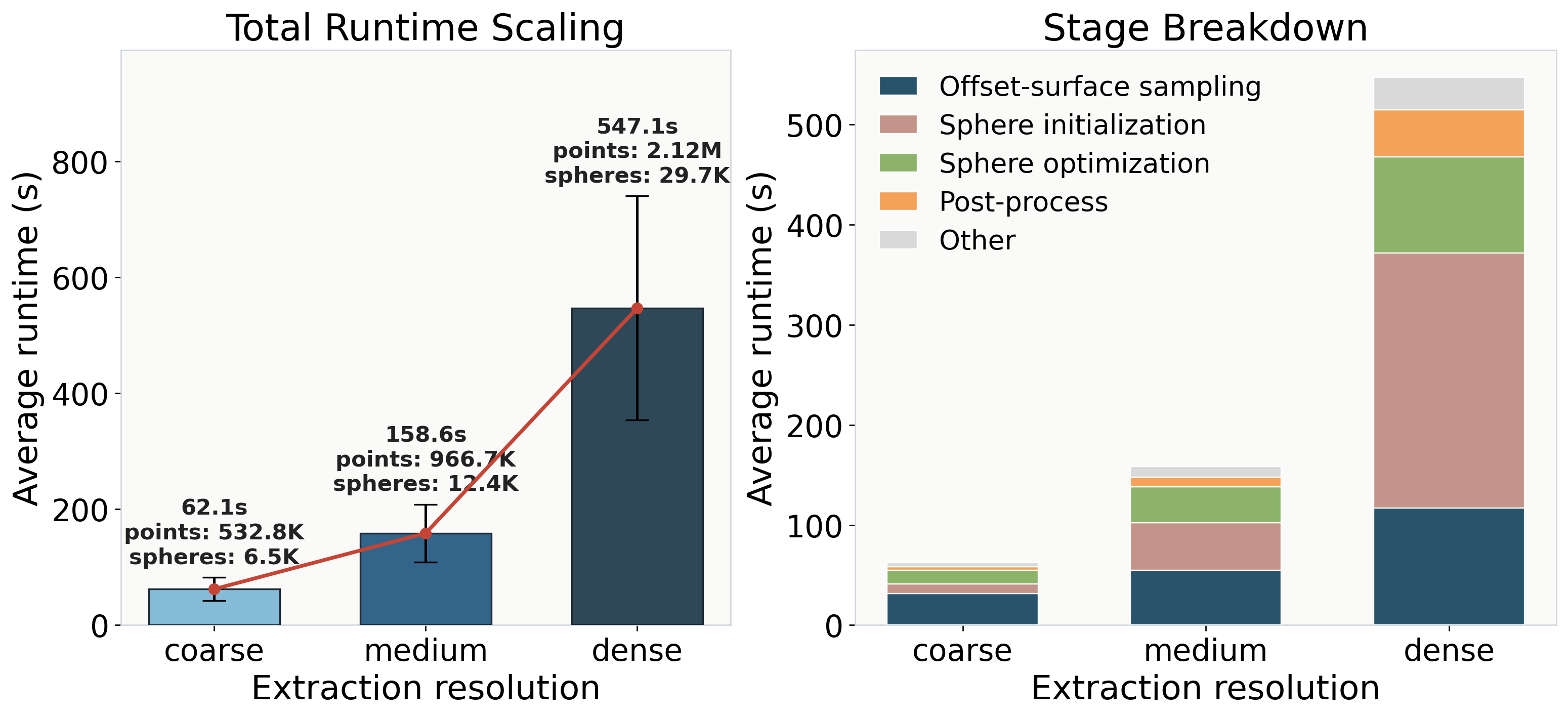}
    \caption{\textbf{Computational time on the DeepFashion dataset:} The left panel reports the mean total runtime under three configurations: coarse, medium, and dense. The right panel shows the average stage-wise runtime.}
    \label{fig:time_scaling}
\end{figure}

Another limitation of our current implementation is its relatively high computational cost, albeit at the benefit of improved reconstruction quality and a more faithful recovery of non-manifold structures.
Figure~\ref{fig:time_scaling} shows that both offset-surface sampling and sphere initialization contribute significantly to the overall runtime, with sphere initialization becoming increasingly expensive for denser reconstructions.
This latter cost mainly stems from the covering-based sphere initialization, which scales roughly as $O(NM)$, with $N$ samples and $M$ selected spheres.
The analyses below further show that the total runtime is strongly correlated with both the number of samples and the number of optimized spheres, with a sharper increase for dense reconstructions.
Exploring alternative sphere initialization strategies is therefore an important direction for future work.

Table~\ref{tab:extraction_time} complements the quantitative comparison in the main text by reporting the average mesh-extraction time per model across datasets and resolutions.

\begin{figure}[h]
    \centering
    \includegraphics[width=0.98\columnwidth]{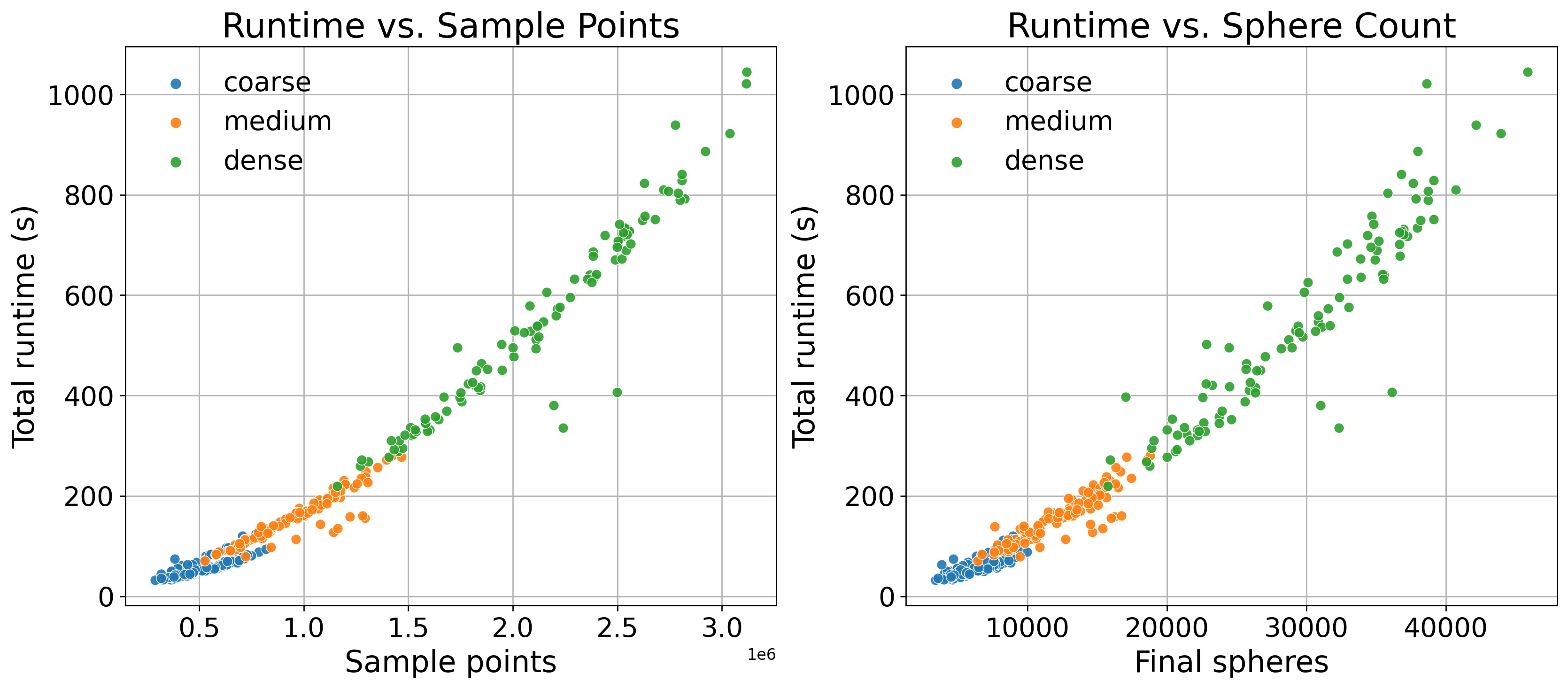}
    \caption{\textbf{Relationship between computational time and problem size:} Left: total runtime versus the number of sample points. Right: total runtime versus the number of final spheres.}
    \label{fig:time_relation}
\end{figure}

Figure~\ref{fig:time_relation} provides additional analysis for the computational-cost limitation discussed in Section~\ref{sec:limitations}. Each point represents one DeepFashion model processed with the coarse, medium, or dense extraction setting. The total runtime increases with both the number of offset-surface samples and the number of final optimized spheres, and this increase becomes more pronounced in the dense regime.

\clearpage
 \begin{figure*}[h]
      \centering
      \includegraphics[width=0.90\linewidth]{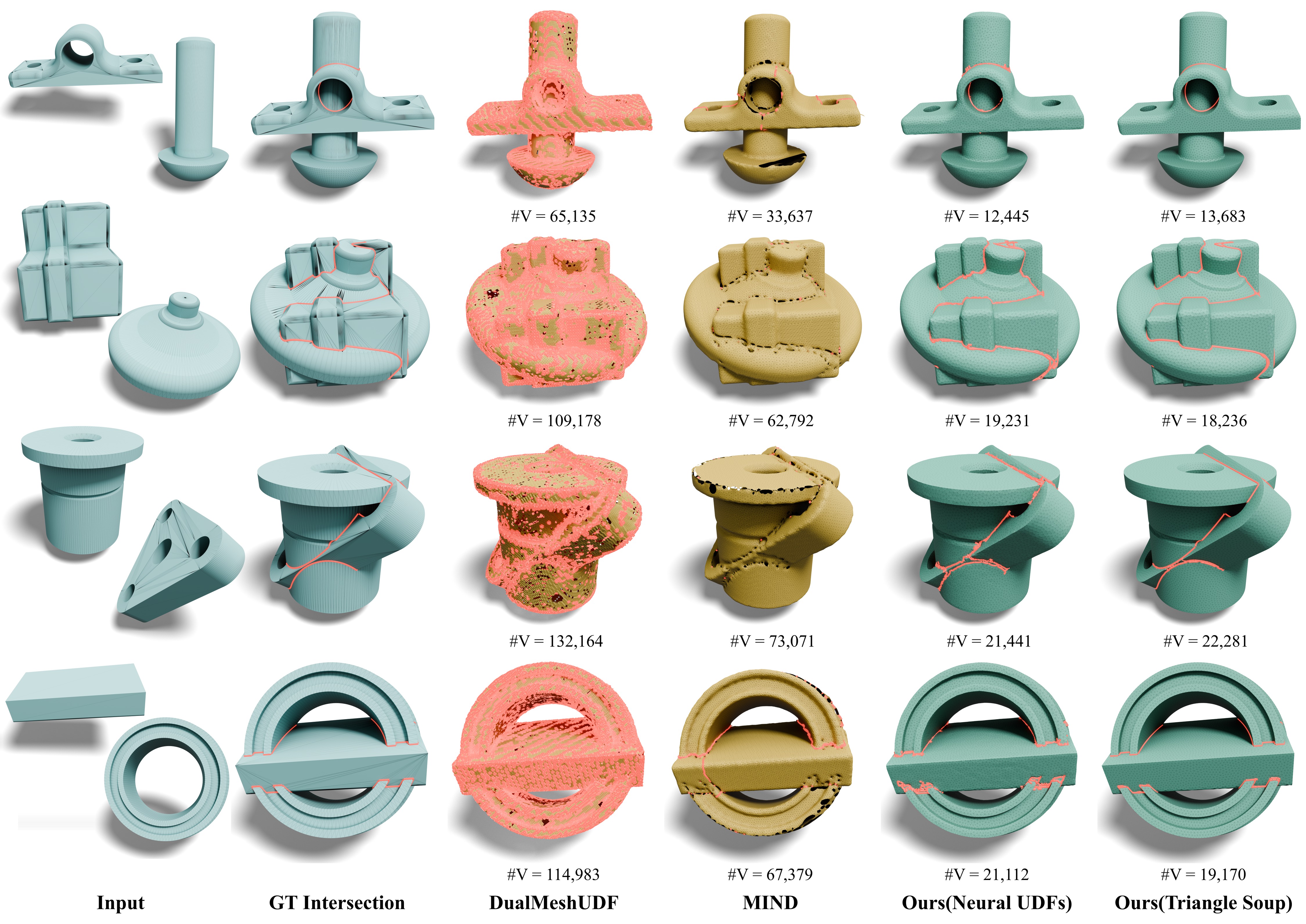}
      \caption{Qualitative comparison of reconstruction results in terms of non-manifold structure preservation.}
      \label{fig:non_manifold}
 \end{figure*}

 \begin{figure*}[h]
      \centering
      \includegraphics[width=0.93\linewidth]{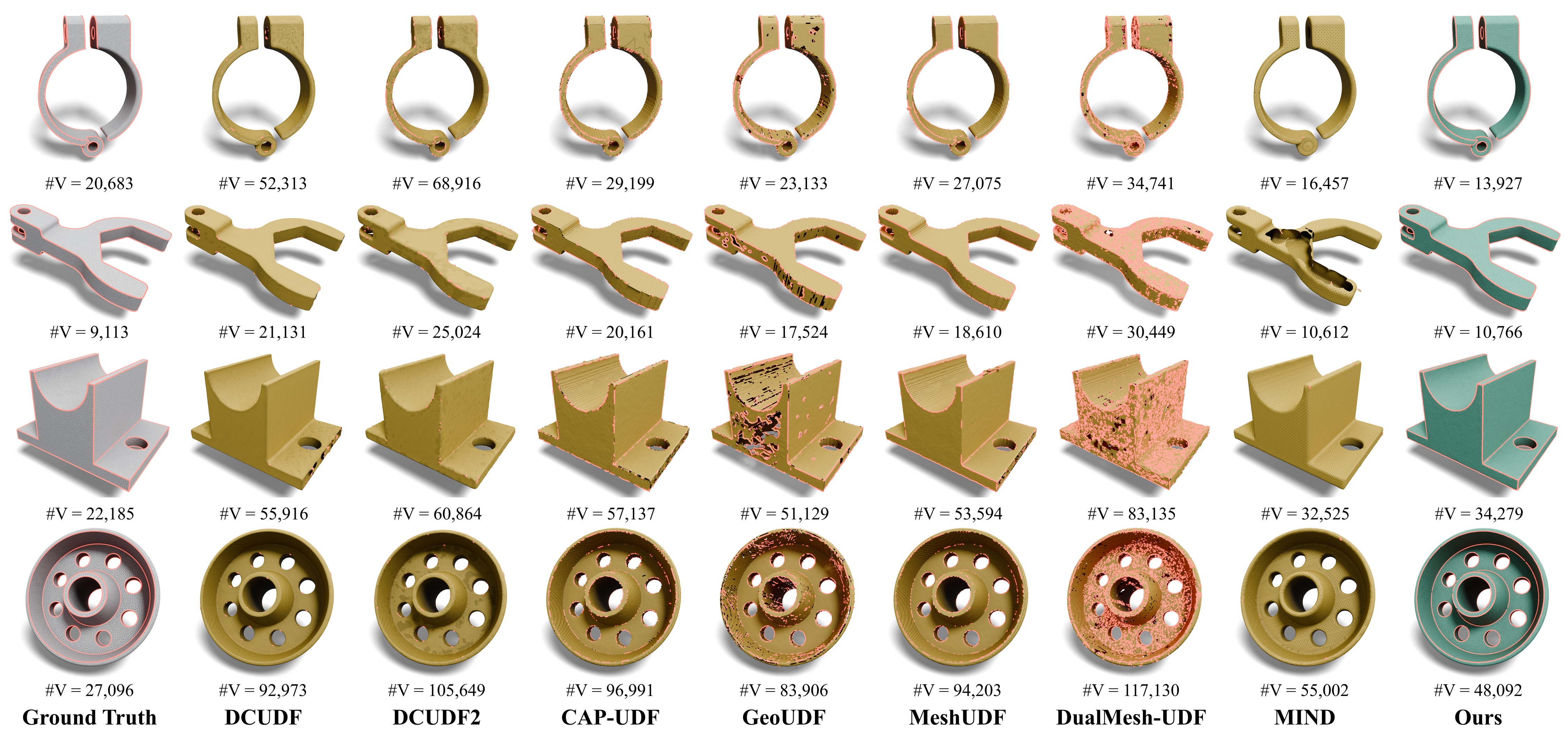}
      \caption{Qualitative comparison of reconstruction results in terms of sharp-edge preservation. A dihedral-angle threshold of $45^\circ$ is used to detect sharp feature edges.}
      \label{fig:sharp_feature}
 \end{figure*}

 \begin{figure*}[p]
     \centering
     \includegraphics[width=0.97\linewidth]{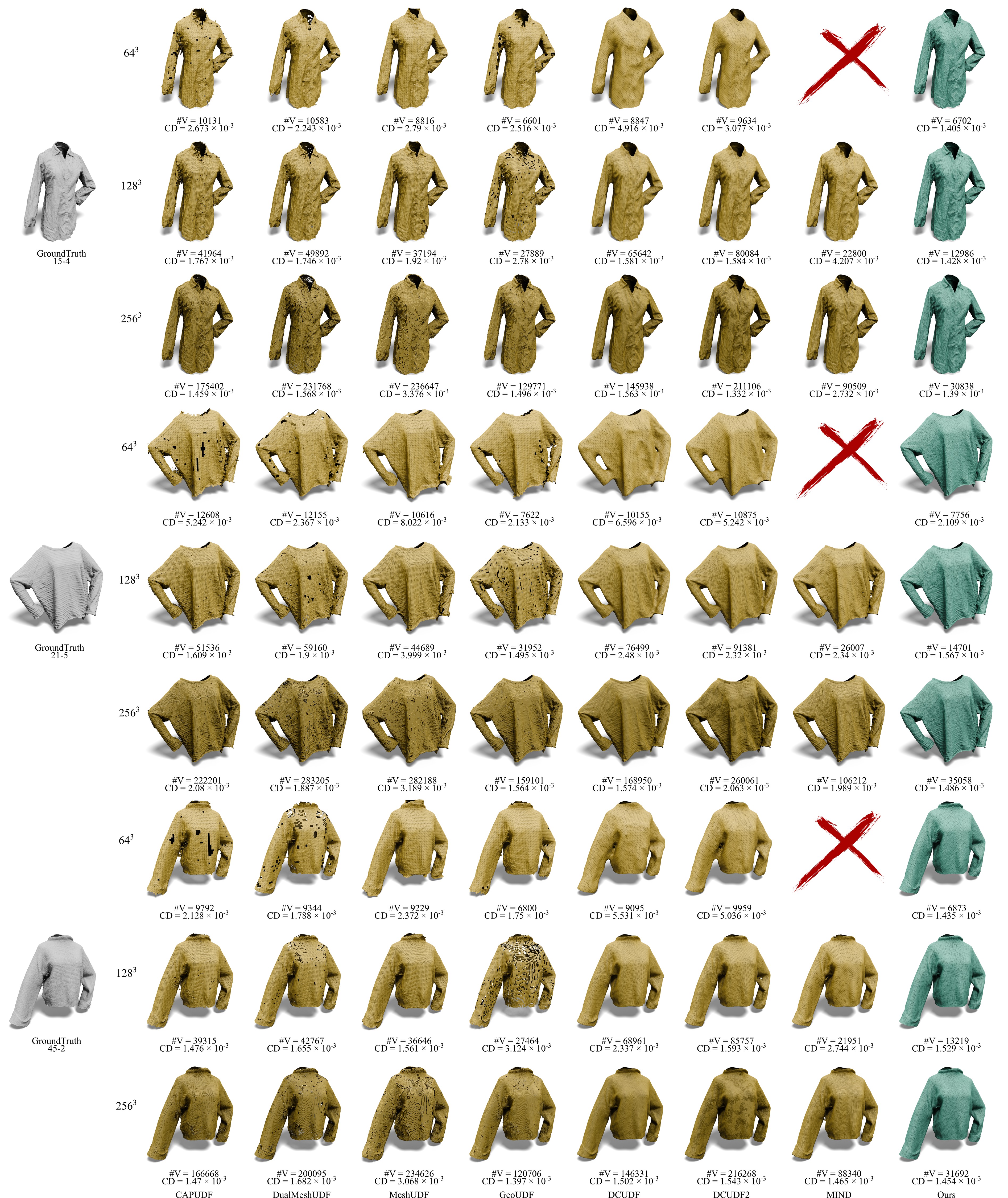}
     \caption{Additional results on the DeepFashion dataset}
     \label{fig:deepfashion}
 \end{figure*}

 \begin{figure*}[p]
     \centering
     \includegraphics[width=\linewidth]{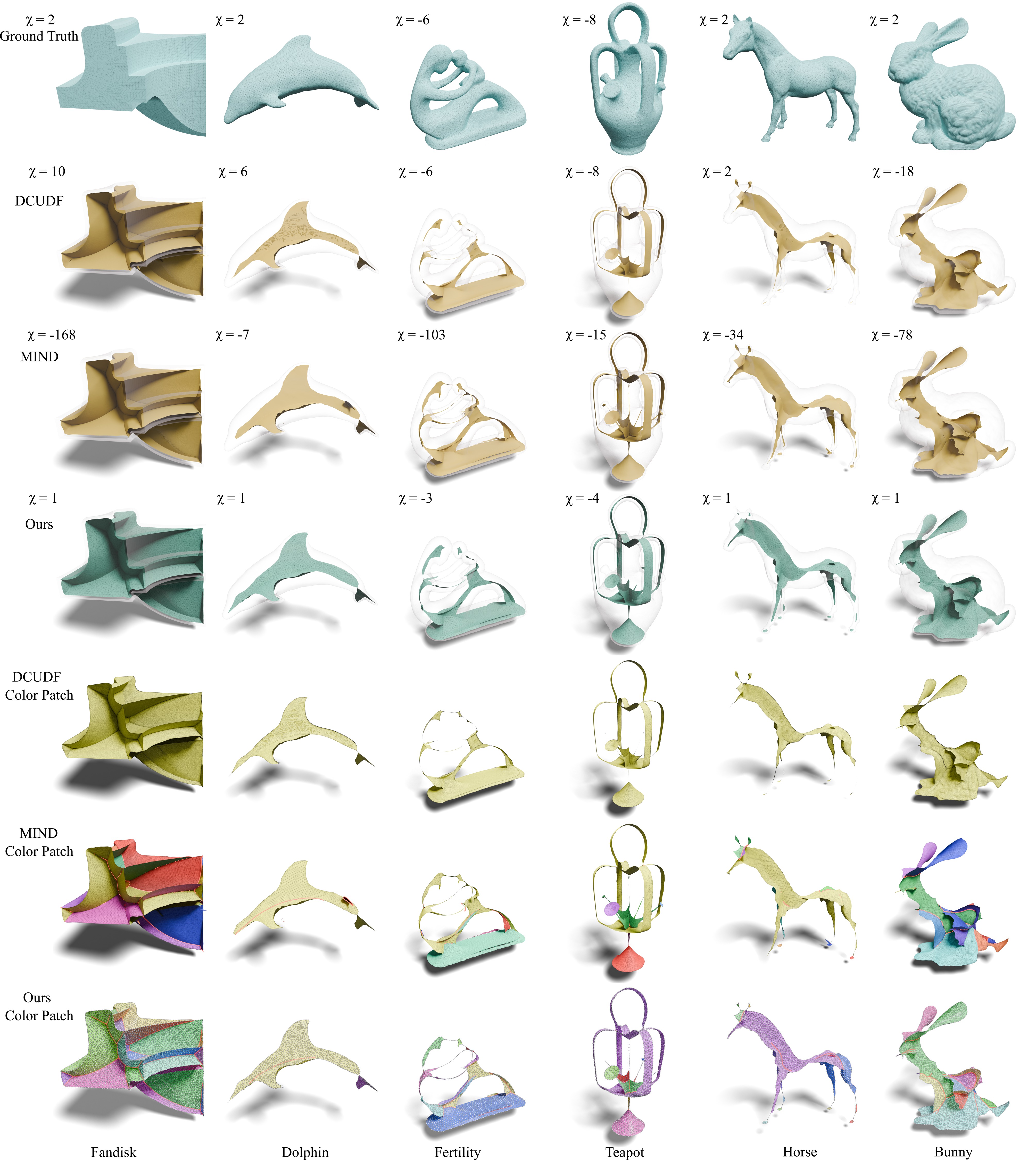}
     \caption{{Qualitative comparison on Q-MDF reconstruction for organic models.} Each column shows one input model and the reconstructions obtained by DCUDF, MIND and our method. The three bottom rows visualize the corresponding manifold patches in different colors, with non-manifold edges highlighted.}
     \label{fig:organic}
 \end{figure*}

 \begin{figure*}[p]
     \centering
     \includegraphics[width=\linewidth]{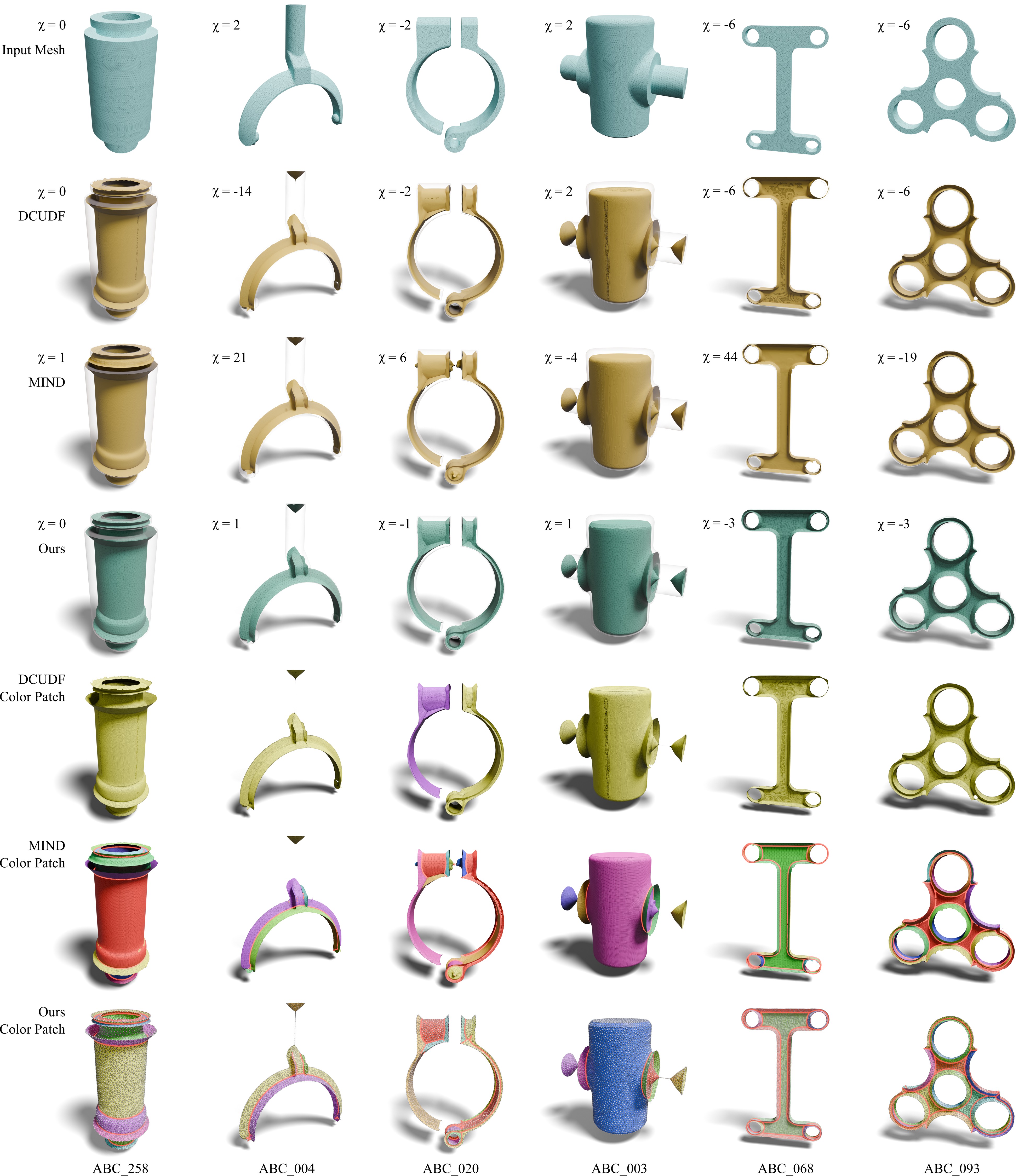}
     \caption{{Qualitative comparison on Q-MDF reconstruction for CAD models.} Each column shows one input model and the reconstructions obtained by DCUDF, MIND and our method. The three bottom rows visualize the corresponding manifold patches in different colors, with non-manifold edges highlighted.}
    \label{fig:ABC}
\end{figure*}

\typeout{get arXiv to do 4 passes: Label(s) may have changed. Rerun}
\end{document}